\documentclass[onecolumn]{aastex631}
\usepackage{float}
\usepackage{multirow}
\usepackage[figuresleft]{rotating}
\newcommand\gr{$\gamma$-ray}
\newcommand\lat{\textit{Fermi}-LAT}

\usepackage{ulem}
\usepackage{xcolor}

\graphicspath{{./}{figures/}}

\begin{document}

\title{GeV Variability Properties of TeV Blazars Detected by \textit{Fermi}-LAT}

\correspondingauthor{Gege Wang}
\email{wanggege@mail.neu.edu.cn}

\correspondingauthor{Junhui Fan}
\email{fjh@gzhu.edu.cn}

\correspondingauthor{Xin Zhang}
\email{zhangxin@mail.neu.edu.cn}

\author{Gege Wang}
\affiliation{Key Laboratory of Cosmology and Astrophysics (Liaoning) $\&$ College of Sciences, Northeastern University, Shenyang 110819, China}

\author{Hubing Xiao}
\affiliation{Shanghai Key Lab for Astrophysics, Shanghai Normal University, Shanghai, 200234, China}

\author{Junhui Fan}
\affiliation{Center for Astrophysics, Guangzhou University, Guangzhou 510006, China}
\affiliation{Astronomy Science and Technology Research Laboratory of Department of Education of Guangdong Province, Guangzhou 510006, China}

\author{Xin Zhang}
\affiliation{Key Laboratory of Cosmology and Astrophysics (Liaoning) $\&$ College of Sciences, Northeastern University, Shenyang 110819, China}
\affiliation{National Frontiers Science Center for Industrial Intelligence and Systems Optimization, Northeastern University, Shenyang 110819, China}
\affiliation{Key Laboratory of Data Analytics and Optimization for Smart Industry (Northeastern University), Ministry of Education, China}

\begin{abstract}

Variability is a prominent observational feature of blazars. 
The high-energy radiation mechanism of jets has always been important but still unclear. 
In this work, we performed a detailed analysis using \lat\ data across 15 years and obtained GeV light curve information for 78 TeV blazars detected by \textit{Fermi}. 
We provided annual GeV fluxes and corresponding spectral indices for the 78 TeV blazars and thorough monthly GeV fluxes for a subsample of 41 bright blazars. 
Our results suggest a strong correlation between the \gr\ photon index and $\log L_{\rm \gamma}$ for the flat spectrum radio quasars (FSRQs) and high-energy-peaked BL Lacs (HBLs). 
14 sources in our sample show significant GeV outbursts/flares above the relatively stable, low-flux light curve, with 6 of them showing a clear sharp peak profile in their 5-day binned light curves. 
We quantified the variability utilizing the fractional variability parameter $F_{\mathrm{var}}$, and found that the flux of the FSRQs showed significantly stronger variability than that of the BL Lacs. 
The 41 bright blazars in this work are best fit by a log-normal flux distribution.
We checked the spectral behavior and found 11 out of the 14 sources show a ‘bluer-when-brighter (BWB)’ trend, suggesting this spectral behavior for these TeV blazars at the GeV band arises from the mechanism that the synchrotron-self Compton (SSC) process dominates the GeV emission. 
Our research offers a systematic analysis of the GeV variability properties of TeV blazars and serves as a helpful resource for further associated blazar studies.

\end{abstract}


\section{Introduction} \label{sec:1}

Blazars are the most powerful active galactic nuclei (AGNs) sources, which show very extreme observational properties, including variability over almost the whole electromagnetic waveband, high and variable polarization, strong \gr\ emissions, and apparent superluminal motion, which are believed to be associated with a relativistic beaming effect of the jet \citep{Urry1995, Villata2006, Fan2014, Gupta2016, Xiao2019, Xiao2022MNRAS, Fan2021}.
Blazars are usually divided into two subclasses: flat spectrum radio quasars (FSRQs) with strong emission lines, and BL Lac objects (BL Lacs) that have weak or even no emission lines \citep{Scarpa1997}.
The broadband emission of blazars ranges from radio band to very-high-energy (VHE), which is generally dominated by non-thermal radiation. 
The broadband spectral energy distribution (SED) of blazar shows two humps, which is generally accepted that the lower energy hump peak is dominated by the synchrotron mechanism. 
The higher energy hump peak could be produced by inverse Compton (IC) scattering of synchrotron photons \citep[synchrotron-self Compton, SSC,][]{bm+96,fin+08} and external photons \citep[external Compton, EC,][]{sik+94,kan+14} in the framework of leptonic models.
Meanwhile, the hadronic model suggests that the higher energy hump is attributed to the proton synchrotron radiation and secondary particle cascade \citep{Mucke2001, Dimitrakoudis2012, Diltz2015, Cerruti2015MNRAS, Xue2021, Wang2022PRD}. 
The hadronic model seems to be promising following the detection of extragalactic neutrino events from the blazar TXS 0506+056 \citep{IceCube2018Sci_1}.

The discovery of the first TeV blazar, Mrk 421, was a surprise when it was detected by the Whipple telescope in 1992 because it was barely seen in the $\gamma$-ray band \citep{Punch1992}.
The following detection of more TeV blazars, e.g., Mrk 501, 1ES 2344+514, PKS 2155-304 \citep{Quinn1996, Catanese1998, Chadwick1999}, started the era of the TeV blazar study.
There are 252 sources associated with TeV emission, and 81 of them are confirmed as blazars according to TeVCat{\footnote{\url{http://tevcat2.uchicago.edu/}}}, the detection of TeV emissions mainly relies on ground-based Cherenkov telescopes, e.g., \textit{Major Atmospheric Gamma Imaging Cherenkov Telescopes} (MAGIC), \textit{High Energy Stereoscopic System telescopes} (H.E.S.S), \textit{Very Energetic Radiation Imaging Telescope Array System} (VERITAS).
Our understanding of blazar TeV emission is limited by several issues, e.g., the sample size of TeV blazars, the lack of TeV light curves due to the observation mode of Cherenkov telescopes, the effective absorption of extragalactic background light (EBL).
A multi-wavelength study is usually employed to investigate the emission properties of TeV blazars, however, this method can only be applied to several individual sources.
Otherwise, we can also study this subject at other bands, for instance, the GeV $\gamma$-ray band.

\textit{Fermi}-LAT, which has been launched since 2008, scans the entire sky every three hours ranging from 20 MeV to above 300 GeV \citep{atw+09}.
During the last 15 years, there are 5 generations of released \textit{Fermi} catalogs with the latest one being the fourth \textit{Fermi}-LAT source catalog (4FGL, \citealp{Abdollahi2020}).
More than 5000 sources have been observed, about 60\% of which are confirmed as blazars and blazars have been established to be the dominant \gr\ sources in the extragalactic sky \citep{3fagn15,4fagn20}.
Based on these observations, people made significant progress in blazar studies, e.g., the classification that depends on the synchrotron peak frequency \citep{Abdo2010ApJ716, Fan2016ApJS226, Yang2022ApJS262}, the `blazar sequence' \citep{Fan2017ApJ835, Ghisellini2017MNRAS469, Ouyang2023ApJ949}, the blazar central engine \citep{Paliya2021ApJS253, Xiao2022ApJ925}.
More studies focus on individual sources, study the properties of flares or outbursts, and put constraints on the blazar emission mechanism, such as the flare of 3C 279 \citep{Shukla2020NatCo11, Tolamatti2022APh13902687, wggpasp}, the neutrino TXS 0506+056 \citep{IceCube2018Sci_2}, variability and spectral properties for 3C 279, Ton 599, and PKS 1222+216\citep{Adams2022ApJ924}, light curve study of PKS 1510+089 \citep{Prince2017ApJ844}.
To obtain information on blazar emission variability, periodicity, and spectrum.
Long-coverage observations on different timescales and spectral analysis can be carried out by taking advantage of the all-sky monitoring capabilities of \textit{Fermi}-LAT. 
Recently, the \textit{Fermi}-LAT light curve repository (LCR), which provides a publicly available, continually updated library of gamma-ray light curves of \textit{Fermi} sources, is released \citep{Abdollahi2023ApJS265}.
However, this library provides light curves binned only on timescales of 3 days, 7 days, and 30 days based on the 10-year \lat\ point source (4FGL-DR2) catalog \citep{4fgldr2}.

In this work, we aim to provide detailed GeV $\gamma$-ray variability information for the TeV blazars based on 15-year 4FGL-DR3 data. 
We described the sample selection and \textit{Fermi} data analysis in Section~\ref{sec:2}. 
The results are reported in Section~\ref{sec:3}. 
The discussion and conclusions are presented in Section~\ref{sec:4} and Section~\ref{sec:5}, respectively. 

\section{Data analysis} \label{sec:2}
\subsection{Sample selection} \label{subsec:2.1}

\begin{figure}
\epsscale{0.8}
\plotone{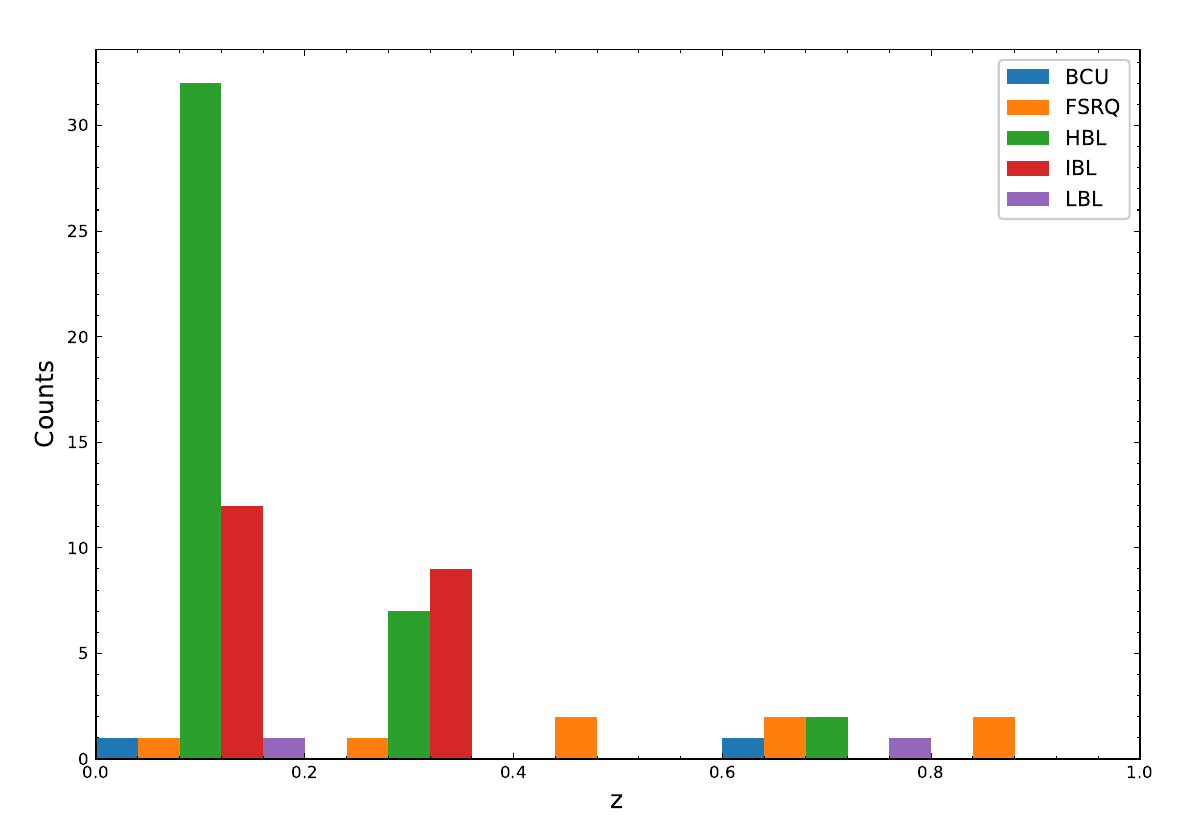}
\caption{The redshift distribution of each type of blazar in the sample.
The histogram is illustrated in 5 bins, which are $0\sim 0.2$, $0.2\sim 0.4$, $0.4\sim 0.6$, $0.6\sim 0.8$, $0.8\sim 1.0$.
The blue bar stands for BCU, the orange bar stands for FSRQ, the green bar stands for HBL, the red bar stands for IBL, and the violet bar stands for LBL.}
\label{fig:z_number}
\end{figure}

We collected 78 blazars, including 66 BL Lacs, 8 FSRQs, and 4 blazar candidates of uncertain type (BCUs) by cross-matching TeVCat and the latest LAT 12-year source (4FGL-DR3) catalog \citep{abd+20}.
These sources are listed in Table \ref{tab:sample}, in which 
columns (1) and (2) give source 4FGL name and associated name;
column (3) gives the redshift obtained from \citet{Chen2018};
column (4) gives the classification that is determined based on the synchrotron peak frequency information and criterion in \citet{Fan2016ApJS226};
We also show the redshift distribution of each type of these blazars in Figure~\ref{fig:z_number}.

\startlongtable
\begin{deluxetable*}{lccccccc}
\label{tab:sample}
\tabletypesize{\scriptsize} \tablecaption{Sample of TeV blazars}
\tablewidth{0pt}
\tablehead{
\colhead{4FGL name} & 
\colhead{Association} &
\colhead{$z$} &
\colhead{Class} &
\colhead{$F_{\gamma}$/$10^{-11}$ (${\rm erg} \cdot {\rm cm}^{-2} \cdot {\rm s}^{-1}$)} &
\colhead{$\Gamma$ ($\alpha$)}  \\
\colhead{(1)} &
\colhead{(2)} &
\colhead{(3)} &
\colhead{(4)} &
\colhead{(5)} &
\colhead{(6)} 
} 
\startdata
J0013.9$-$1854	&	SHBL	J001355.9$-$185406	&	0.094	&	IBL		&	0.2	$\pm$	0.03	&	2.05	$\pm$	0.13		\\
J0033.5$-$1921	&	KUV	00311$-$1938	&	0.61	&	HBL		&	222.4	$\pm$	11.1	&	1.47	$\pm$	0.07		\\
J0035.9+5950	&	1ES	0033+595	&	0.086	&	HBL		&	 56.6	$\pm$	 1.8	&		1.53 $\pm$		0.02 	\\
J0112.1+2245	&	S2	0109+22	&	0.265	&	IBL		&	43.6	$\pm$	2.2	&	1.96	$\pm$	0.07	&		\\
J0136.5+3906	&	RGB	J0136+391	&	0.75	&	HBL		&	318.9	$\pm$	9.8	&	1.34	$\pm$	0.05		\\
J0152.6+0147	&	RGB	J0152+017	&	0.08	&	IBL		&	0.7	$\pm$	0.06	&	1.96	$\pm$	0.07		\\
J0214.3+5145	&	TXS	0210+515	&	0.049	&	HBL		&	0.5	$\pm$	0.05	&	1.9	$\pm$	0.09		\\
J0221.1+3556	&	S3	0218+35	&	0.944	&	FSRQ		&		5.9 $\pm$	 1.1	&	 2.30	$\pm$		0.01 	\\
J0222.6+4302	&	3C	66A	&	0.34	&	IBL		&	62.9	$\pm$	1.5	&	1.84	$\pm$	0.04		\\
J0232.8+2018	&	1ES	0229+200	&	0.139	&	HBL		&	0.4	$\pm$	0.06	&	1.74	$\pm$	0.11		\\
J0238.4$-$3116	&	1RXS	J023832.6$-$311658	&	0.232	&	HBL		&	1.3	$\pm$	0.08	&	1.79	$\pm$	0.05		\\
J0303.4$-$2407	&	PKS	0301$-$243	&	0.26	&	IBL		&	10.8	$\pm$	0.6	&	1.86	$\pm$	0.07		\\
J0319.8+1845	&	RBS	413	&	0.19	&	HBL		&	0.8	$\pm$	0.08	&	1.72	$\pm$	0.08		\\
J0349.4$-$1159	&	1ES	0347$-$121	&	0.188	&	HBL		&	0.6	$\pm$	0.06	&	1.73	$\pm$	0.09		\\
J0416.9+0105	&	1ES	0414+009	&	0.287	&	HBL		&	0.7	$\pm$	0.07	&	1.8	$\pm$	0.08		\\
J0449.4$-$4350	&	PKS	0447$-$439	&	0.205	&	HBL		&	169	$\pm$	3.4	&	1.67	$\pm$	0.03		\\
J0507.9+6737	&	1ES	0502+675	&	0.34	&	HBL		&	3.9	$\pm$	0.04	&	1.54	$\pm$	0.01		\\
J0509.4+0542	&	TXS	0506+056	&	0.337	&	IBL		&	29.6	$\pm$	1.1	&	1.98	$\pm$	0.06		\\
J0521.7+2112	&	VER	J0521+211	&	0.108	&	HBL		&	30.7	$\pm$	0.7	&	1.87	$\pm$	0.04		\\
J0550.5$-$3216	&	PKS	0548$-$322	&	0.069	&	HBL		&	0.4	$\pm$	0.05	&	1.86	$\pm$	0.1		\\
J0648.7+1516	&	RX	J0648.7+1516	&	0.179	&	HBL		&	2	$\pm$	0.05	&	1.73	$\pm$	0.02		\\
J0650.7+2503	&	1ES	0647+250	&	0.203	&	HBL		&	30.1	$\pm$	1	&	1.59	$\pm$	0.04		\\
J0710.4+5908	&	RGB	J0710+591	&	0.125	&	HBL		&	0.9	$\pm$	0.08	&	1.62	$\pm$	0.06		\\
J0721.9+7120	&	S5	0716+714	&	0.3	&	IBL		&	326.3	$\pm$	8.7	&	1.85	$\pm$	0.03		\\
J0733.4+5152	&	PGC	2402248	&	0.065	&	BCU		&	0.4	$\pm$	0.02	&	1.69	$\pm$	0.02		\\
J0739.2+0137	&	PKS	0736+017	&	0.189	&	FSRQ		&	35.8	$\pm$	4	&	2.26	$\pm$	0.12		\\
J0809.8+5218	&	1ES	0806+524	&	0.138	&	HBL		&	15	$\pm$	0.6	&	1.75	$\pm$	0.06		\\
J0812.0+0237	&	1RXS	J081201.8+023735	&	0.172	&	HBL		&	0.6	$\pm$	0.03	&	1.83	$\pm$	0.03		\\
J0847.2+1134	&	RBS	723	&	0.199	&	HBL		&	0.7	$\pm$	0.07	&	1.7	$\pm$	0.08		\\
J0854.8+2006	&	OJ	287	&	0.306	&	IBL		&	6.9	$\pm$	0.5	&	2.18	$\pm$	0.1		\\
J0904.9$-$5734	&	PKS	0903$-$57	&	0.695	&	BCU		&	7	$\pm$	1.7	&	2.07	$\pm$	0.03		\\
J0958.7+6534	&	S4	0954+65	&	0.367	&	IBL		&	15	$\pm$	1.2	&	2.08	$\pm$	0.1		\\
J1010.2$-$3119	&	1RXS	J101015.9$-$311909	&	0.143	&	HBL		&	0.9	$\pm$	0.08	&	1.75	$\pm$	0.07		\\
J1015.0+4926	&	1ES	1011+496	&	0.212	&	HBL		&	33.4	$\pm$	1.1	&	1.76	$\pm$	0.05		\\
J1058.6+2817	&	GB6	J1058+2817	&	/	&	IBL		&	0.7	$\pm$	0.02	&	2.14	$\pm$	0.02		\\
J1103.6$-$2329	&	1ES	1101$-$232	&	0.186	&	HBL		&	0.8	$\pm$	0.09	&	1.63	$\pm$	0.08		\\
J1104.4+3812	&	Mrk	421	&	0.031	&	HBL		&	139.2	$\pm$	1.9	&	1.67	$\pm$	0.02		\\
J1136.4+6736	&	RX	J1136.5+6737	&	0.136	&	HBL		&	0.7	$\pm$	0.06	&	1.73	$\pm$	0.06		\\
J1136.4+7009	&	Mrk	180	&	0.045	&	HBL		&	1.4	$\pm$	0.08	&	1.79	$\pm$	0.04		\\
J1159.5+2914	&	TON	599	&	0.729	&	FSRQ		&	18.7	$\pm$	1.3	&	2.19	$\pm$	0.08		\\
J1217.9+3007	&	1ES	1215+303	&	0.13	&	IBL		&	54.5	$\pm$	1.7	&	1.8	$\pm$	0.05		\\
J1221.3+3010	&	1ES	1218+304	&	0.182	&	HBL		&	6.3	$\pm$	0.2	&	1.7	$\pm$	0.02		\\
J1221.5+2814	&	W	Comae	&	0.103	&	IBL		&	2.1	$\pm$	0.09	&	2.19	$\pm$	0.04		\\
J1224.4+2436	&	MS	1221.8+2452	&	0.219	&	IBL		&	1.1	$\pm$	0.07	&	1.95	$\pm$	0.05		\\
J1224.9+2122	&	4C	21.35	&	0.435	&	FSRQ		&	31.2	$\pm$	2.5	&	2.23	$\pm$	0.08		\\
J1230.2+2517	&	S3	1227+25	&	0.135	&	IBL		&	32.3	$\pm$	2.2	&	1.95	$\pm$	0.09		\\
J1256.1$-$0547	&	3C	279	&	0.536	&	FSRQ		&	456.1	$\pm$	20.3	&	2.1	$\pm$	0.05		\\
J1315.0$-$4236	&	1ES	1312$-$423	&	0.105	&	HBL		&	0.7	$\pm$	0.08	&	1.69	$\pm$	0.09		\\
J1422.3+3223	&	B2	1420+32	&	0.682	&	FSRQ		&	122.2	$\pm$	0.4	&	1.94	$\pm$	0.02		\\
J1427.0+2348	&	PKS	1424+240	&	0.16	&	IBL		&	196.3	$\pm$	5.3	&	1.62	$\pm$	0.04		\\
J1428.5+4240	&	H	1426+428	&	0.129	&	HBL		&	1.3	$\pm$	0.09	&	1.62	$\pm$	0.05		\\
J1442.7+1200	&	1ES	1440+122	&	0.163	&	HBL		&	0.8	$\pm$	0.08	&	1.7	$\pm$	0.07		\\
J1443.9+2501	&	PKS	1441+25	&	0.939	&	FSRQ		&	241.8	$\pm$	17	&	1.85	$\pm$	0.1		\\
J1443.9$-$3908	&	PKS	1440$-$389	&	0.065	&	HBL		&	60.9	$\pm$	2.1	&	1.65	$\pm$	0.05		\\
J1512.8$-$0906	&	PKS	1510$-$089	&	0.36	&	FSRQ		&	35.6	$\pm$	1	&	2.38	$\pm$	0.04		\\
J1517.7$-$2422	&	AP	Lib	&	0.048	&	IBL		&	13.8	$\pm$	0.5	&	2.01	$\pm$	0.02		\\
J1518.0$-$2731	&	TXS	1515$-$273	&	0.128	&	LBL		&	1.4	$\pm$	0.08	&	2.05	$\pm$	0.05		\\
J1555.7+1111	&	PG	1553+113	&	0.36	&	HBL		&	721.2	$\pm$	12.6	&	1.43	$\pm$	0.03		\\
J1653.8+3945	&	Mrk	501	&	0.034	&	HBL		&	22.2	$\pm$	0.5	&	1.74	$\pm$	0.04		\\
J1725.0+1152	&	H	1722+119	&	0.018	&	HBL		&	82.5	$\pm$	3	&	1.65	$\pm$	0.05		\\
J1728.3+5013	&	1ES	1727+502	&	0.055	&	HBL		&	2.3	$\pm$	0.1	&	1.77	$\pm$	0.03		\\
J1744.0+1935	&	1ES	1741+196	&	0.084	&	HBL		&	0.8	$\pm$	0.06	&	1.93	$\pm$	0.06		\\
J1751.5+0938	&	OT	81	&	0.322	&	LBL		&	20.2	$\pm$	1.4	&	2.13	$\pm$	0.09		\\
J1813.5+3144	&	B2	1811+31	&	0.117	&	IBL		&	0.5	$\pm$	0.3	&	1.93	$\pm$	0.08		\\
J1857.9+0313c	&	MAGIC	J1857.6+0297	&	/	&	BCU		&	0.6	$\pm$	0.1	&	3.15	$\pm$	0.29		\\
J1944.0+2117	&	HESS	J1943+213	&	/	&	HBL		&	2.2	$\pm$	0.3	&	1.34	$\pm$	0.14		\\
J2000.0+6508	&	1ES	1959+650	&	0.047	&	HBL		&	28.1	$\pm$	0.6	&	1.74	$\pm$	0.03		\\
J2001.2+4353	&	MAGIC	J2001+435	&	0.174	&	IBL		&	49.9	$\pm$	2	&	1.78	$\pm$	0.07		\\
J2009.4$-$4849	&	PKS	2005$-$489	&	0.071	&	HBL		&	5.6	$\pm$	0.3	&	1.82	$\pm$	0.09		\\
J2039.5+5218	&	1ES	2037+521	&	0.053	&	IBL		&	0.7	$\pm$	0.08	&	1.77	$\pm$	0.09		\\
J2056.7+4939	&	RGB	J2056+496	&	/	&	BCU		&	2.2	$\pm$	0.04	&	1.85	$\pm$	0.01		\\
J2158.8$-$3013	&	PKS	2155$-$304	&	0.116	&	HBL	&	116.6	$\pm$	2.4	&	1.72	$\pm$	0.03		\\
J2202.7+4216	&	BL	Lacertae	&	0.069	&	IBL		&	72.2	$\pm$	1.8	&	2.11	$\pm$	0.04		\\
J2243.9+2021	&	RGB	J2243+203	&	0.39	&	IBL		&	65.5	$\pm$	2.5	&	1.71	$\pm$	0.06		\\
J2250.0+3825	&	B3	2247+381	&	0.119	&	IBL		&	1	$\pm$	0.08	&	1.72	$\pm$	0.06		\\
J2324.7$-$4041	&	1ES	2322$-$409	&	0.174	&	HBL		&	11.8	$\pm$	0.8	&	1.61	$\pm$	0.09		\\
J2347.0+5141	&	1ES	2344+514	&	0.044	&	HBL		&	3.3	$\pm$	0.1	&	1.88	$\pm$	0.04		\\
J2359.0$-$3038	&	H	2356$-$309	&	0.165	&	HBL		&	0.6	$\pm$	0.06	&	1.79	$\pm$	0.08		\\
\enddata
\tablecomments{Here we use the classification reported in \citet{Fan2016ApJS226}. 
Low-energy-peaked BL Lacs (LBL): 
for BL Lacs with the synchrotron-peak frequency $\log \nu_{\mathrm{p}}(\mathrm{Hz}) \leqslant$ 14.0; intermediate-energy-peaked BL Lacs (IBL): $14.0 < \log \nu_{\mathrm{p}}(\mathrm{Hz}) \leqslant$ 15.3; high-energy-peaked BL Lacs (HBL): $\log \nu_{\mathrm{p}}(\mathrm{Hz}) > 15.3$. 
$F_{\gamma}$ and $\Gamma^{\rm ph}_{\gamma}$ are the 1--300 GeV energy flux and photon index of the maximum likelihood analysis results over 15 years, respectively.} 
\end{deluxetable*}

\subsection{\textit{Fermi}-LAT observations and data reduction} \label{subsec:2.2}

LAT is one of the main instruments on board \textit{Fermi}. 
LAT scans the whole sky every three hours in the energy range from 20 MeV to $>$300 GeV \citep{atw+09}. We selected LAT data from the Fermi Pass 8 database in the time period from 2008 August 4 15:43:36
(UTC) to 2023 Mar 9 03:03:00 (UTC), with an energy range of 1--300 GeV. 
Following the recommendations of the LAT team\footnote{\footnotesize http://fermi.gsfc.nasa.gov/ssc/data/analysis/scitools/}, we selected events with zenith angles less than 90 deg to prevent possible contamination from the Earth's limb. 
The LAT science tool Fermitools 2.0.8 and the instrument response function (IRF) P8R3\_SOURCE\_V2 were used. 
For the selected samples, a 20$^\circ\times 20^\circ$ square region of interest (ROI) centered at their positions given in 4FGL-DR3 was selected. 
The normalization parameters and spectral indices of the sources within 5 deg from the target, as well as sources within the ROI with variable index $\geq$ 72.44 \citep{ace+15}, were set as free parameters. 
All other parameters were fixed at their catalog values in 4FGL-DR3. 
We used the original spectral models in 4FGL-DR3 for the sources in the source model when performing binned maximum likelihood analysis with \texttt{gtlike}. 
A simple power-law ($d N / d E \propto E^{-\Gamma}$, where $\Gamma$ is the photon index) spectral type was used for each blazar when deriving its light curve. 
We checked through the likelihood analysis results assuming a power-law model comparing with a log-parabola ($dN/dE \propto (E/E_0)^{-\alpha - \beta\log (E/E_0)}$, where $\alpha$ and $\beta$ are spectral parameters) for the samples in the source models, and found that a log-parabola is not significantly preferred over a power-law for the samples, except for J0035.9+5950 and J0221.1+3556.
Therefore we changed the two sources' spectral parameters accordingly in Table~\ref{tab:sample}.
The comparison was conducted by calculating $\sqrt{-2\log(L_{pl}/L_{logP})}$, where $L_{pl}$ and $L_{logP}$ are the maximum likelihood values obtained from a power-law and a log-parabola, respectively \citep{abd+13}.
In addition, the background galactic and extragalactic diffuse emission models were added to the source model using the spectral model file gll\_iem\_v07.fits and iso\_P8R3\_SOURCE\_V2\_v1.txt, respectively. 
The normalizations of the two diffuse emission components were set as free parameters in the analysis.
We constructed light curves binned in 90-day time intervals by performing standard binned maximum likelihood analysis, calculated flux ($F_{\rm \gamma}$) and photon spectral index ($\Gamma$) for the energy range of 1--300 GeV spectrum and listed them in columns (5) and (6) of Table ~\ref{tab:sample}.

\section{Results} \label{sec:3}
\subsection{Annual and monthly intensity at the GeV band} \label{sec:3.1}
\lat\ has conducted observations at \gr\ energy bands over 15 years by scanning the whole sky every three hours. 
As we aim to provide detailed GeV spectral behaviors, study the GeV variability of the TeV blazars, and put constraints on the blazar emission model.
We calculated the annual GeV fluxes and corresponding photon spectral indices for the 78 TeV blazars in our sample and listed them in Table \ref{tab:annual_lc}, in which the annual (360-day time interval) maximum, the minimum, and the mean fluxes and the corresponding photon spectral indices are given for the past 15 years since the launch of \textit{Fermi} (MJD 54683).

\begin{figure}
\centering
\includegraphics[width=0.3\textwidth]{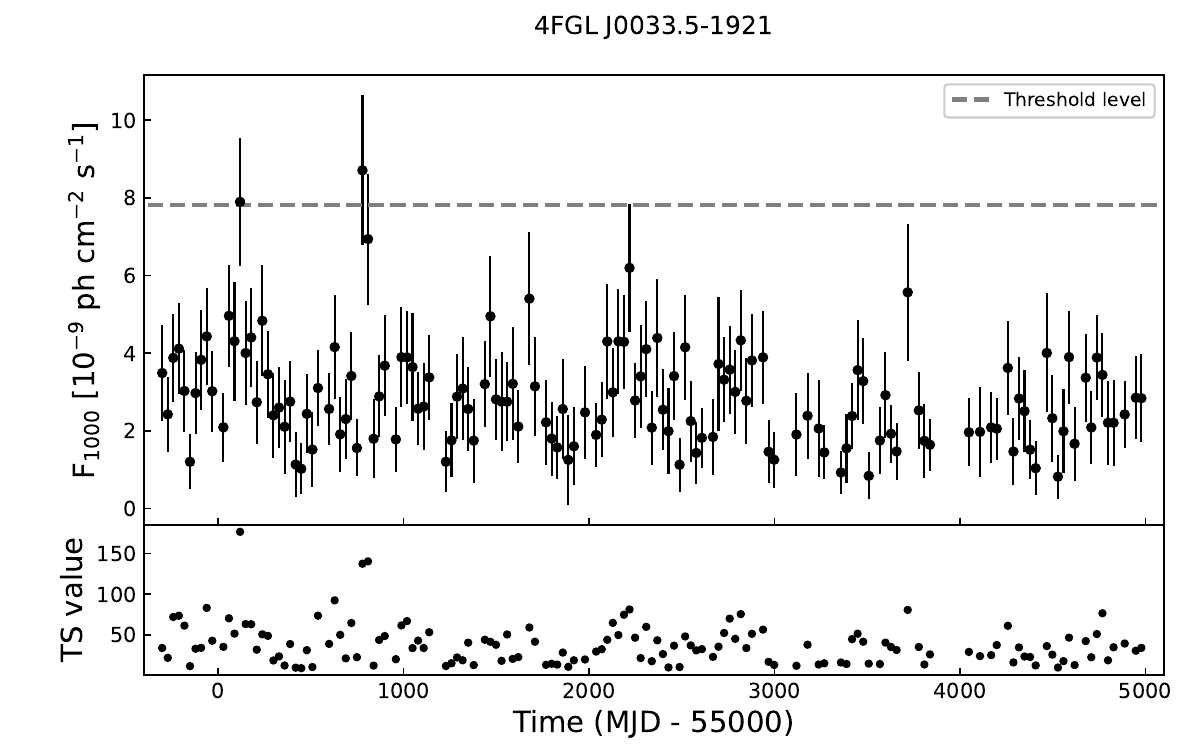}
\includegraphics[width=0.3\textwidth]{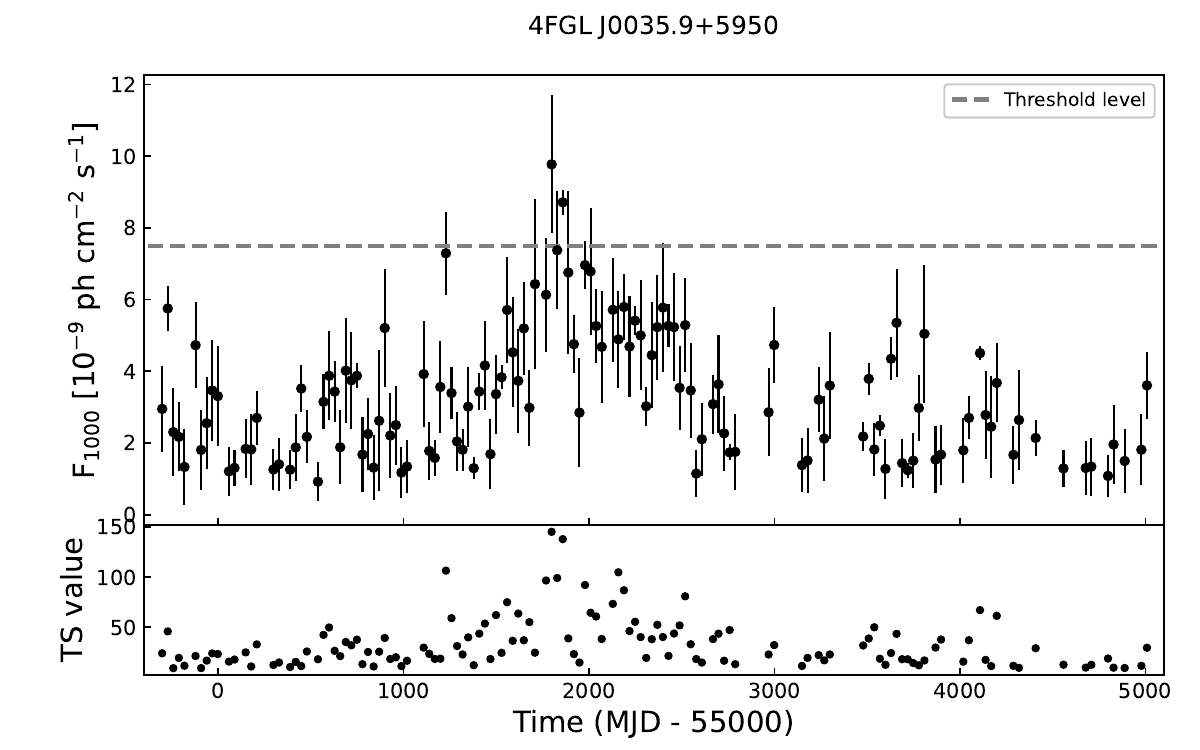}
\includegraphics[width=0.3\textwidth]{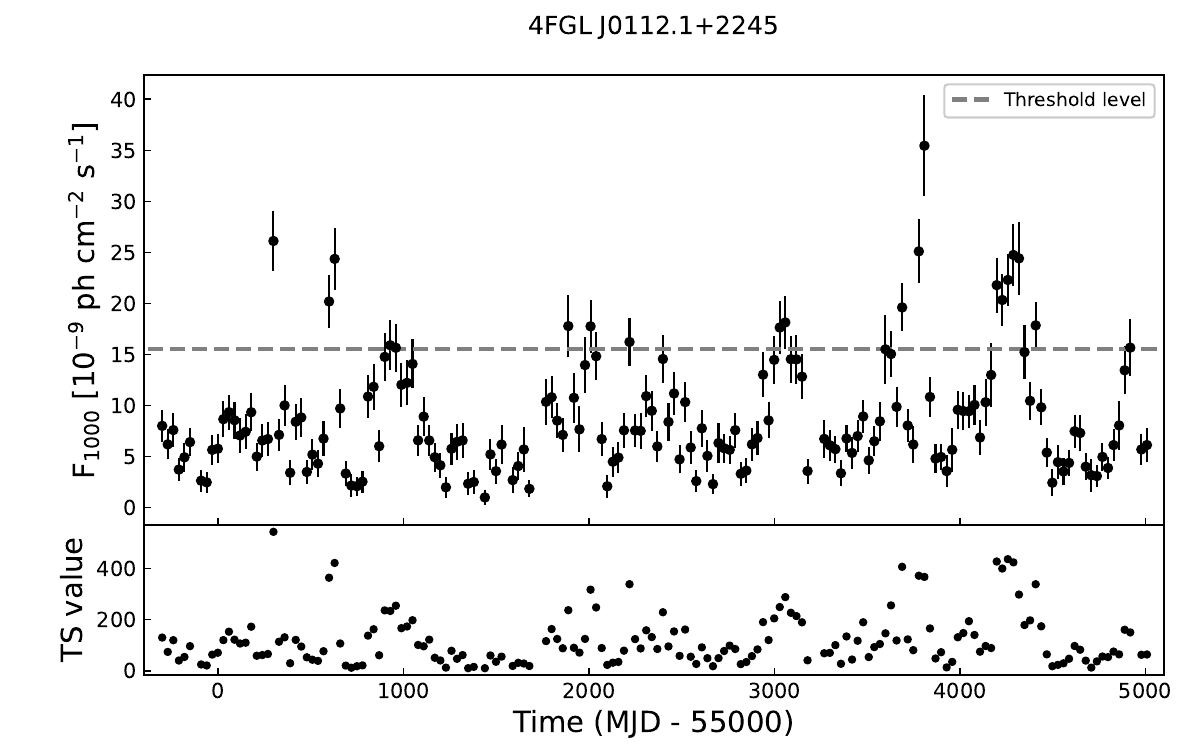}

\includegraphics[width=0.3\textwidth]{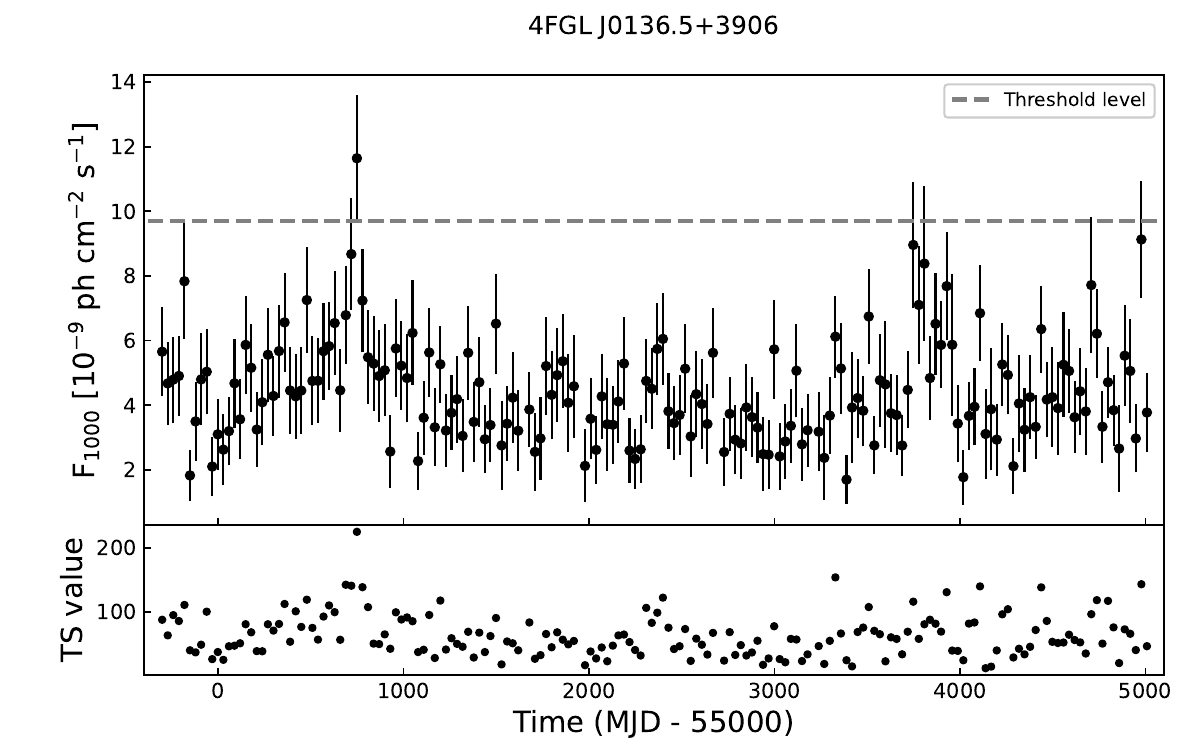}
\includegraphics[width=0.3\textwidth]{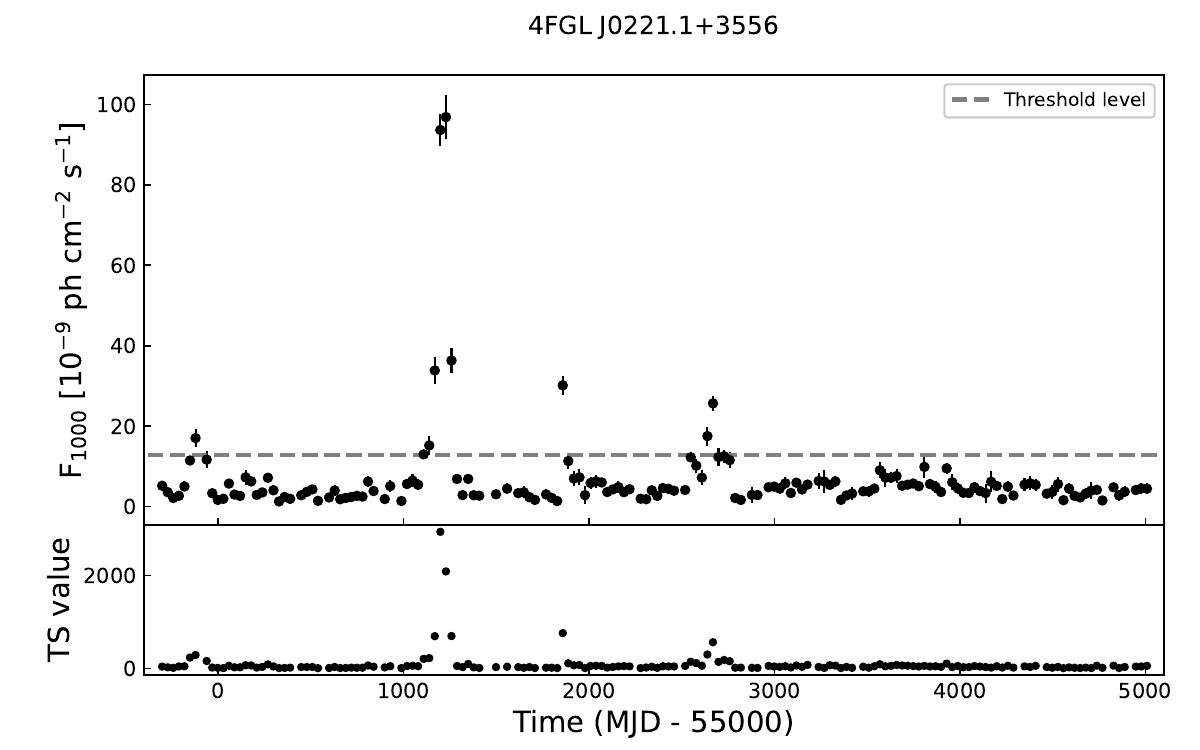}
\includegraphics[width=0.3\textwidth]{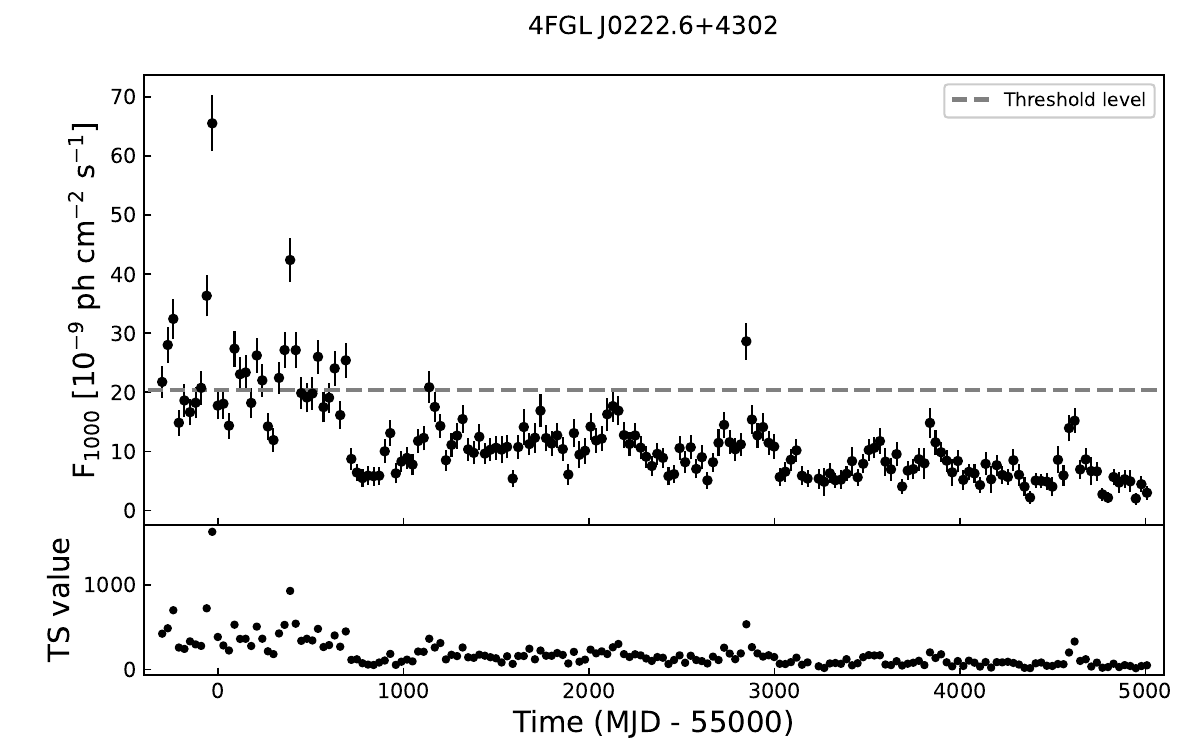}
\caption{The monthly binned light curves for the 41 bright blazars in
    our sample. Only six items are presented here. The complete figure set 
    (41 images) is available in the online journal.}
\label{fig:30dlc}
\end{figure}

Moreover, according to the 90-day binned light curves we select a subsample of bright blazars.
The selection criterion is that the source has at least one-third of data points that have the maximum likelihood Test Statistic (TS) values larger than 75 (three times the 5$\sigma$ detection significance), there are 41 blazars are selected and marked by `Y' of the Bright flag in Table~\ref{tab:annual_lc}. 
We further constructed 30-day binned light curves for these 41 TeV blazars. 
The monthly binned light curves are shown in Figure~\ref{fig:30dlc}, for which only the flux data points with the maximum likelihood TS values being larger than 9 are plotted. 
To further investigate the spectral behavior of these 41 bright TeV blazars, we calculated the detailed monthly flux and corresponding photon spectral index and listed in Table~\ref{tab:monthly_lc_flux} and Table~\ref{tab:monthly_lc_index}.
In these two tables, the MJD time represents the beginning of each bin. 
The TS values for each MJD time of each blazar are listed in parentheses. 
Note that in some periods, there may be situations where the spectral photon index is too large or too small, which requires simultaneous consideration of the TS value. 
Usually, we use data points with TS values larger than 9 in the analysis.

\begin{sidewaystable}
  \caption{\mbox{Annual GeV fluxes and photon indices for the 78 TeV blazars of our sample.}}
  \label{tab:annual_lc}
  \fontsize{7.5}{7.5}\selectfont
\centering
  \renewcommand{\arraystretch}{1.2}
  \begin{tabular}{llcccccccccc}
\hline
\qquad & Name  & Bright  & 2008-2009 &  2009-2010 &  2010-2011  & 2011-2012 & 2012-2013 & 2013-2014 & 2014-2015 &  ... & 2022-2023  \\ 
\hline
$f_{\rm \gamma}^{\rm max}$   & J0013.9$-$1854  &   & 7.2  $\pm$  3.5  & 5.5  $\pm$  3.5  & 12  $\pm$  4.7  & 5.6  $\pm$  3.4  & 8.7  $\pm$  4.1  & 3.9  $\pm$  3.4  & 5.7  $\pm$  3.2  &  ...  & 5.2  $\pm$  11  \\
$f_{\rm \gamma}^{\rm min}$   &   &   & 0.00082  $\pm$  1.1  & 1.2  $\pm$  1.2  & 0.0067  $\pm$  0.55  & 2  $\pm$  2.6  & 6.2e-05  $\pm$  0.00023  & 1.2  $\pm$  1.9  & 0.67  $\pm$  2.2  &  ...  & 9.5e-05  $\pm$  0.15  \\
$f_{\rm \gamma}^{\rm mean}$   &   &   & 3.3  & 2.4  & 4  & 3.6  & 3.3  & 2.5  & 3.4  &  ...  & 1.7  \\
$\Gamma^{\rm max}$   &   &   & 2.6  $\pm$  1  & 2.8  $\pm$  1.9  & 10  $\pm$  0.068  & 5.8  $\pm$  3.5  & 2.1  $\pm$  1.6e+02  & 10  $\pm$  1.3  & 3.3  $\pm$  3.2  &  ...  & 4.2  $\pm$  11  \\
$\Gamma^{\rm min}$   &   &   & 1.7  $\pm$  0.55  & 0.78  $\pm$  0.66  & 2.6  $\pm$  0.59  & 2.5  $\pm$  1.2  & 1.4  $\pm$  0.6  & 1.1  $\pm$  0.6  & 1.4  $\pm$  0.34  &  ...  & 1.3  $\pm$  7.5e+02  \\
$\Gamma^{\rm mean}$   &   &   & 2  & 1.8  & 5.4  & 4.3  & 1.8  & 3.9  & 2.3  &  ...  & 2.8  \\
\hline
$f_{\rm \gamma}^{\rm max}$   & J0033.5$-$1921  &  Y & 50  $\pm$  13  & 79  $\pm$  17  & 87  $\pm$  19  & 69  $\pm$  17  & 49  $\pm$  16  & 54  $\pm$  17  & 62  $\pm$  16  &  ...  & 34  $\pm$  11  \\
$f_{\rm \gamma}^{\rm min}$   &   &   & 10  $\pm$  6.6  & 11  $\pm$  8.3  & 9.9  $\pm$  7.4  & 11  $\pm$  8  & 7.5  $\pm$  7.4  & 16  $\pm$  8  & 12  $\pm$  12  &  ...  & 22  $\pm$  11  \\
$f_{\rm \gamma}^{\rm mean}$   &   &   & 31  & 36  & 28  & 32  & 24  & 26  & 32  &  ...  & 27  \\
$\Gamma^{\rm max}$   &   &   & 2.7  $\pm$  0.63  & 2.6  $\pm$  0.55  & 10  $\pm$  0.068  & 2.8  $\pm$  0.78  & 5.3  $\pm$  3.7  & 3.4  $\pm$  1.1  & 2.3  $\pm$  2.1  &  ...  & 2.3  $\pm$  0.4  \\
$\Gamma^{\rm min}$   &   &   & 1.5  $\pm$  0.29  & 1.5  $\pm$  0.24  & 0.9  $\pm$  0.35  & 1.5  $\pm$  0.25  & 1.5  $\pm$  0.29  & 0.77  $\pm$  0.63  & 1.3  $\pm$  0.39  &  ...  & 1.1  $\pm$  0.23  \\
$\Gamma^{\rm mean}$   &   &   & 2  & 2  & 2.4  & 1.8  & 2.5  & 2  & 1.7  &  ...  & 1.8  \\
\hline
$f_{\rm \gamma}^{\rm max}$   &  J0035.9+5950 &  Y & 58  $\pm$  6.3  & 27  $\pm$  7.5  & 40  $\pm$  15  & 52  $\pm$  17  & 73  $\pm$  11  & 98  $\pm$  19  & 70  $\pm$  6.6  &  ...  & 36  $\pm$  9.3  \\
$f_{\rm \gamma}^{\rm min}$   &   &   & 4.7  $\pm$  3.9  & 8.9  $\pm$  4.7  & 4.8  $\pm$  4.7  & 12  $\pm$  7  & 13  $\pm$  3  & 9  $\pm$  7.6  & 28  $\pm$  15  &  ...  & 4.5  $\pm$  4.4  \\
$f_{\rm \gamma}^{\rm mean}$   &   &   & 26  & 16  & 27  & 24  & 31  & 54  & 53  &  ...  & 15  \\
$\Gamma^{\rm max}$   &   &   & 5  $\pm$  0.001  & 5  $\pm$  0.73  & 3.7  $\pm$  0.0029  & 3.2  $\pm$  0.0054  & 2.7  $\pm$  0.19  & 2.2  $\pm$  0.34  & 2.7  $\pm$  0.053  &  ...  & 2.9  $\pm$  0.0032  \\
$\Gamma^{\rm min}$   &   &   & 0.15  $\pm$  0.12  & 0.14  $\pm$  0.12  & 0.82  $\pm$  0.024  & 0.65  $\pm$  0.25  & 0.32  $\pm$  0.014  & 0.32  $\pm$  0.15  & 0.56  $\pm$  0.049  &  ...  & 0.92  $\pm$  0.14  \\
$\Gamma^{\rm mean}$   &   &   & 2  & 2.2  & 2.1  & 1.8  & 1.4  & 1.4  & 1.7  &  ...  & 1.9  \\
\hline
$f_{\rm \gamma}^{\rm max}$   & J0112.1+2245  & Y  & 93  $\pm$  17  & 2.6e+02  $\pm$  29  & 2.4e+02  $\pm$  30  & 1.6e+02  $\pm$  24  & 66  $\pm$  17  & 1.1e+02  $\pm$  21  & 1.8e+02  $\pm$  31  &  ...  & 1.6e+02  $\pm$  28  \\
$f_{\rm \gamma}^{\rm min}$   &   &   & 16  $\pm$  10  & 34  $\pm$  12  & 21  $\pm$  9.3  & 60  $\pm$  16  & 9.8  $\pm$  6.9  & 9.1  $\pm$  6  & 21  $\pm$  11  &  ...  & 39  $\pm$  11  \\
$f_{\rm \gamma}^{\rm mean}$   &   &   & 56  & 88  & 77  & 1.1e+02  & 39  & 52  & 1e+02  &  ...  & 80  \\
$\Gamma^{\rm max}$   &   &   & 4.5  $\pm$  1.7  & 2.6  $\pm$  0.41  & 3.8  $\pm$  1.1  & 2.9  $\pm$  0.38  & 6.7  $\pm$  2.9  & 3  $\pm$  1.1  & 3  $\pm$  0.76  &  ...  & 2.6  $\pm$  0.29  \\
$\Gamma^{\rm min}$   &   &   & 1.9  $\pm$  0.37  & 1.8  $\pm$  0.2  & 1.7  $\pm$  0.27  & 1.8  $\pm$  0.2  & 1.9  $\pm$  0.58  & 1.5  $\pm$  0.47  & 1.5  $\pm$  0.35  &  ...  & 2  $\pm$  0.23  \\
$\Gamma^{\rm mean}$   &   &   & 2.6  & 2.1  & 2.5  & 2.2  & 2.8  & 2.4  & 2.1  &  ...  & 2.3  \\
\hline
$f_{\rm \gamma}^{\rm max}$   & J0136.5+3906  & Y  & 78  $\pm$  18  & 66  $\pm$  15  & 1.2e+02  $\pm$  19  & 62  $\pm$  16  & 65  $\pm$  16  & 54  $\pm$  15  & 53  $\pm$  14  &  ...  & 91  $\pm$  18  \\
$f_{\rm \gamma}^{\rm min}$   &   &   & 18  $\pm$  7.9  & 32  $\pm$  12  & 45  $\pm$  13  & 23  $\pm$  8.9  & 29  $\pm$  10  & 14  $\pm$  11  & 16  $\pm$  13  &  ...  & 27  $\pm$  13  \\
$f_{\rm \gamma}^{\rm mean}$   &   &   & 42  & 48  & 65  & 47  & 41  & 37  & 35  &  ...  & 46  \\
$\Gamma^{\rm max}$   &   &   & 2.1  $\pm$  0.3  & 2.2  $\pm$  0.36  & 2.2  $\pm$  0.31  & 2.4  $\pm$  0.38  & 2.2  $\pm$  0.39  & 2.1  $\pm$  0.28  & 2.2  $\pm$  0.97  &  ...  & 2  $\pm$  0.28  \\
$\Gamma^{\rm min}$   &   &   & 1.4  $\pm$  0.29  & 1.3  $\pm$  0.2  & 1.5  $\pm$  0.16  & 1.4  $\pm$  0.3  & 1.5  $\pm$  0.24  & 1.5  $\pm$  0.21  & 1.4  $\pm$  0.26  &  ...  & 1.4  $\pm$  0.33  \\
$\Gamma^{\rm mean}$   &   &   & 1.8  & 1.8  & 1.7  & 1.7  & 1.8  & 1.7  & 1.8  &  ...  & 1.7  \\
\hline
...   & ...  & ...  & ...  & ...  & ...  & ... & ... & ... & ...  & ... & ...   \\
\hline 
\end{tabular} \\
\begin{minipage}[]{200mm}
{Notes: $f_{\rm \gamma}$ in units of $10^{-10}$ ${\rm ph} \cdot {\rm cm}^{-2} \cdot {\rm s}^{-1}$}, and only five items are presented.
\end{minipage}
\end{sidewaystable}

\begin{sidewaystable}
  \caption{\mbox{Monthly GeV photon fluxes of the 41 bright TeV blazars}}
  \label{tab:monthly_lc_flux}
  \fontsize{7.5}{7.5}\selectfont
\centering
  \renewcommand{\arraystretch}{1.2}
  \begin{tabular}{ccccccccc}
\hline
MJD  & J0033.5$-$1921 & J0035.9+5950 & J0112.1+2245 & J0136.5+3906 & J0221.1+3556 & J0222.6+4302  & J0303.4$-$2407 & ... \\ 
\hline
54683	&	3.49e+01	$\pm$	1.24e-09	&	2.95e-09	$\pm$	1.21e-09	&	8.00e-09	$\pm$	1.56e-09	&	5.66e-09	$\pm$	1.38e-09	&	5.18e-09	$\pm$	8.65e-10	&	2.18e-08	$\pm$	2.68e-09	&	2.97e-09	$\pm$	1.01e-09 & ...\\
54713	&	2.42e-09	$\pm$	9.60e-10	&	5.75e-09	$\pm$	6.32e-10	&	6.19e-09	$\pm$	1.46e-09	&	4.67e-09	$\pm$	1.28e-09	&	3.57e-09	$\pm$	9.36e-10	&	2.80e-08	$\pm$	3.06e-09	&	3.40e-09	$\pm$	1.09e-09 & ...\\
54743	&	3.88e-09	$\pm$	1.13e-09	&	2.30e-09	$\pm$	1.22e-09	&	7.59e-09	$\pm$	1.64e-09	&	4.79e-09	$\pm$	1.33e-09	&	2.13e-09	$\pm$	1.03e-09	&	3.24e-08	$\pm$	3.43e-09	&	4.62e-09	$\pm$	1.31e-09 & ...\\
54773	&	4.11e-09	$\pm$	1.17e-09	&	2.17e-09	$\pm$	9.61e-10	&	3.71e-09	$\pm$	1.12e-09	&	4.91e-09	$\pm$	1.24e-09	&	2.68e-09	$\pm$	9.27e-10	&	1.48e-08	$\pm$	2.18e-09	&	2.11e-09	$\pm$	9.25e-10 & ...\\
54803	&	3.02e-09	$\pm$	1.05e-09	&	1.33e-09	$\pm$	1.05e-09	&	4.92e-09	$\pm$	1.40e-09	&	7.84e-09	$\pm$	1.79e-09	&	4.99e-09	$\pm$	1.24e-09	&	1.86e-08	$\pm$	2.78e-09	&	3.10e-09	$\pm$	1.09e-09 & ...\\
... & ... & ... & ... & ... & ... & ... & ...  & ...\\
\hline 
\end{tabular} \\
\begin{minipage}[]{200mm}
{Notes: $f_{\rm \gamma}$ in units of ${\rm ph} \cdot {\rm cm}^{-2} \cdot {\rm s}^{-1}$.
Table \ref{tab:monthly_lc_flux} is published in its entirety in the machine-readable format. A portion is shown here for guidance regarding its form and content.}
\end{minipage}
\end{sidewaystable}

\begin{sidewaystable}
  \caption{\mbox{Monthly GeV photon indices of the 41 bright TeV blazars}}
  \label{tab:monthly_lc_index}
  \fontsize{7.5}{7.5}\selectfont
\centering
  \renewcommand{\arraystretch}{1.2}
  \begin{tabular}{ccccccccc}
\hline
MJD  & J0033.5$-$1921 & J0035.9+5950 & J0112.1+2245 & J0136.5+3906 & J0221.1+3556 & J0222.6+4302  & J0303.4$-$2407 & ... \\ 
\hline
54683 	&	2.06 	$\pm$	0.34 	(33.87) 	&	1.48 	$\pm$	0.39 	(23.96) 	&	2.29 	$\pm$	0.24 	(129.66) 	&	1.81 	$\pm$	0.21 	(87.97) 	&	3.15 	$\pm$	0.03 	(44.27) 	&	1.89 	$\pm$	0.11 	(422.97) 	&	1.99 	$\pm$	0.34 	(40.08)   & ...\\ 
54713 	&	2.68 	$\pm$	0.63 	(21.77) 	&	1.52 	$\pm$	0.07 	(45.82) 	&	2.75 	$\pm$	0.38 	(73.61) 	&	1.93 	$\pm$	0.25 	(63.71) 	&	4.39 	$\pm$	0.04 	(28.20) 	&	2.14 	$\pm$	0.12 	(486.64) 	&	2.19 	$\pm$	0.37 	(53.06)   & ...\\ 
54743 	&	1.71 	$\pm$	0.23 	(72.02) 	&	2.05 	$\pm$	0.29 	(9.12) 	&	2.13 	$\pm$	0.23 	(119.35) 	&	1.46 	$\pm$	0.20 	(95.25) 	&	3.88 	$\pm$	0.03 	(17.48) 	&	1.77 	$\pm$	0.09 	(699.83) 	&	2.06 	$\pm$	0.29 	(65.38)    & ...\\
54773 	&	1.71 	$\pm$	0.23 	(73.58) 	&	0.96 	$\pm$	0.57 	(19.40)	&	2.89 	$\pm$	0.52 	(39.78) 	&	1.75 	$\pm$	0.20 	(86.03) 	&	0.72 	$\pm$	1.37 	(46.57) 	&	2.21 	$\pm$	0.17 	(258.71) 	&	1.94 	$\pm$	0.47 	(25.62)    & ...\\
54803 	&	1.46 	$\pm$	0.24 	(61.40) 	&	1.03 	$\pm$	0.06 	(11.54) 	&	2.49 	$\pm$	0.39 	(53.79) 	&	1.90 	$\pm$	0.21 	(111.16) 	&	2.43 	$\pm$	0.09 	(50.73) 	&	2.38 	$\pm$	0.20 	(242.92)	&	1.75 	$\pm$	0.29 	(47.64)    & ...\\
... & ... & ... & ... & ... & ... & ... & ...  & ...\\
\hline 
\end{tabular} \\
\begin{minipage}[]{200mm}
{Table \ref{tab:monthly_lc_index} is published in its entirety in the machine-readable format. A portion is shown here for guidance regarding its form and content.}
\end{minipage}
\end{sidewaystable}

\subsection{The GeV luminosity and spectral photon index} \label{subsec:3.2}
\begin{figure*}
\centering
\includegraphics[width=0.7\textwidth]{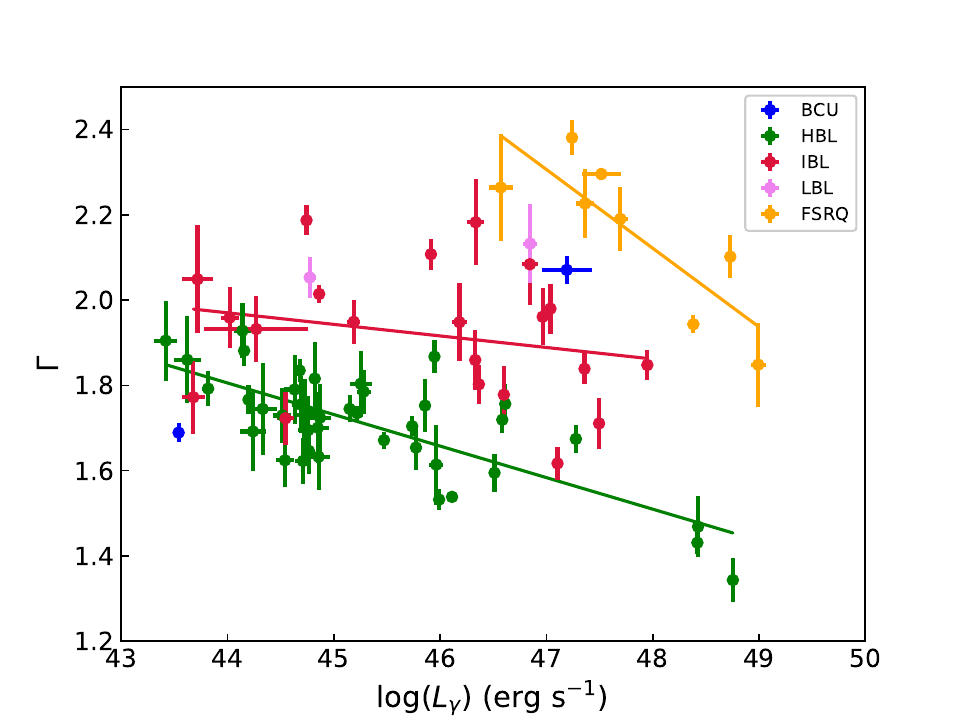}
\caption{The correlation between 1--300 GeV photon index and luminosity of 74 blazars. The solid lines are the linear regressions fitting results.}
\label{fig:luminosity}
\end{figure*}

The $\gamma$-ray luminosity is calculated by
\begin{equation}
L_{\rm \gamma} = 4\pi d_{\rm L}^2(1+z)^{(\Gamma -2)}F_{\rm \gamma},
\end{equation}
where 
\begin{equation}
d_{\rm L} = \frac{c}{H_{\rm 0}}\int^{1+z}_{1}\frac{1}{\sqrt{\Omega_{\rm m}x^{3}+1-\Omega_{\rm m}}}dx 
\end{equation}
is a luminosity distance \citep{Komatsu2011} and $(1+z)^{(\Gamma - 2)}$ stands for a $K$-correction.
We calculated the \gr\ luminosity of the 74 blazars that have redshift information, using the energy flux derived from the binned likelihood analysis, and studied the correlations between the GeV $\gamma$-ray luminosity and photon index in Figure \ref{fig:luminosity}.

It is found that FSRQs occupy the upper-right region, then the IBLs occupy the middle region, and the HBLs occupy the lower-left region of Figure \ref{fig:luminosity}.
This result suggests that the TeV blazars show a decrease in the GeV $\gamma$-ray luminosity and photon spectral index with the increase of synchrotron peak frequency, and indicates a `blazar sequence' that was initially proposed by \citet{Fossati1998}.
In addition, we calculated the linear regressions between $\Gamma$ and $\log L_\gamma$ as $$\Gamma =(-0.18 \pm 0.05)\log L_\gamma +(10.98 \pm 2.36)$$
with the correlation coefficient $r=-0.84$ and the chance probability $p$=0.01 for FSRQs through Pearson analysis;
$$\Gamma =(-0.03 \pm 0.03) \log L_\gamma +(3.16 \pm 1.19)$$
with $r=-0.23$, $p$=0.31 for IBLs;
$$\Gamma =(-0.07 \pm 0.01)\log L_\gamma +(5.06 \pm 0.48)$$
with $r=-0.74$, $p=2.60 \times 10^{-8}$ for HBLs.
The regression results are shown in Figure \ref{fig:luminosity} and suggest a strong correlation between $\Gamma$ and $\log L_{\rm \gamma}$ for the FSRQs and HBLs, while no correlation for IBLs. 
We also conducted statistics on weighted Kendall's tau \citep{SHIEH199817} and Spearman's coefficients. 
The $r$ value obtained through weighted Kendall's tau analysis was $-$0.74, $-$0.60, $-$0.28 for FSRQs, HBLs, and IBLs respectively. 
The coefficient obtained using Spearman's statistics is $r=-0.86$, $p=6.53 \times 10^{-3}$, $r=-0.59$, $p=4.96 \times 10^{-5}$ and $r=-0.26$, $p=0.26$ for FSRQs, HBLs and IBLs respectively. 
These results have supported that of Pearson's analysis mentioned above.

\citet{3lac+15,4fagn20} showed the LAT photon index versus the \gr\ luminosity for the different blazar subclasses of the whole sample in the Third LAT AGN Catalog (3LAC) and the Fourth LAT AGN Catalog (4LAC) blazars. 
The trend of softer spectra with higher luminosity reported in previous catalogs is also confirmed. 
However, 4LAC noted that the correlation between photon index and \gr\ luminosity is significant overall for blazars, but much weaker when the different classes are taken independently. 
While 3LAC \gr\ luminosity results were computed from the 4-year \lat\ point source (3FGL) catalog energy flux between 100 MeV and 100 GeV. 
They also mentioned that due to the bias in the selection criteria for the 57 BL Lacs with both lower and upper limits on their redshifts or only upper limits, the HBLs with both limits were found to be more luminous on average than those with measured redshifts.

\subsection{GeV flux and spectral photon index in flares} \label{subsec:3.3}
\begin{figure*}
\centering
\includegraphics[width=0.31\textwidth]{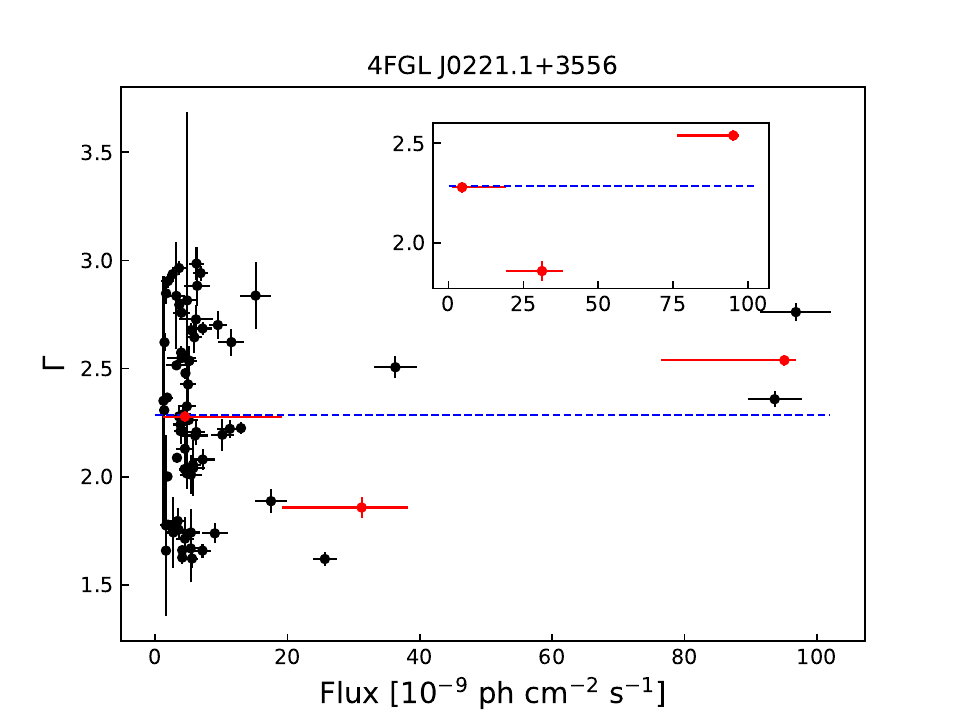}
\includegraphics[width=0.31\textwidth]{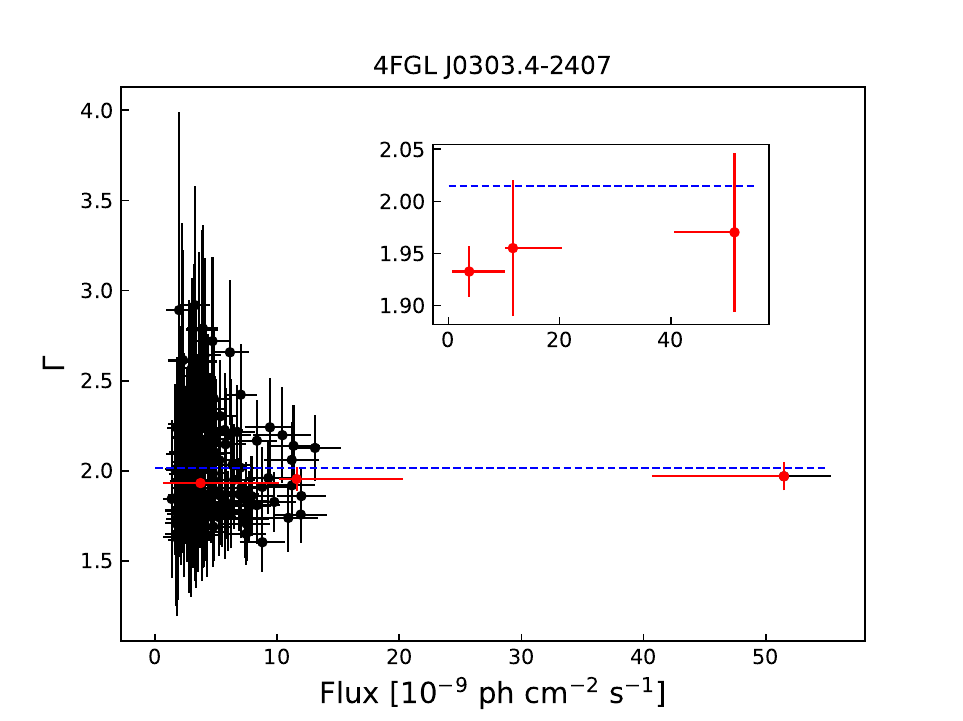}
\includegraphics[width=0.31\textwidth]{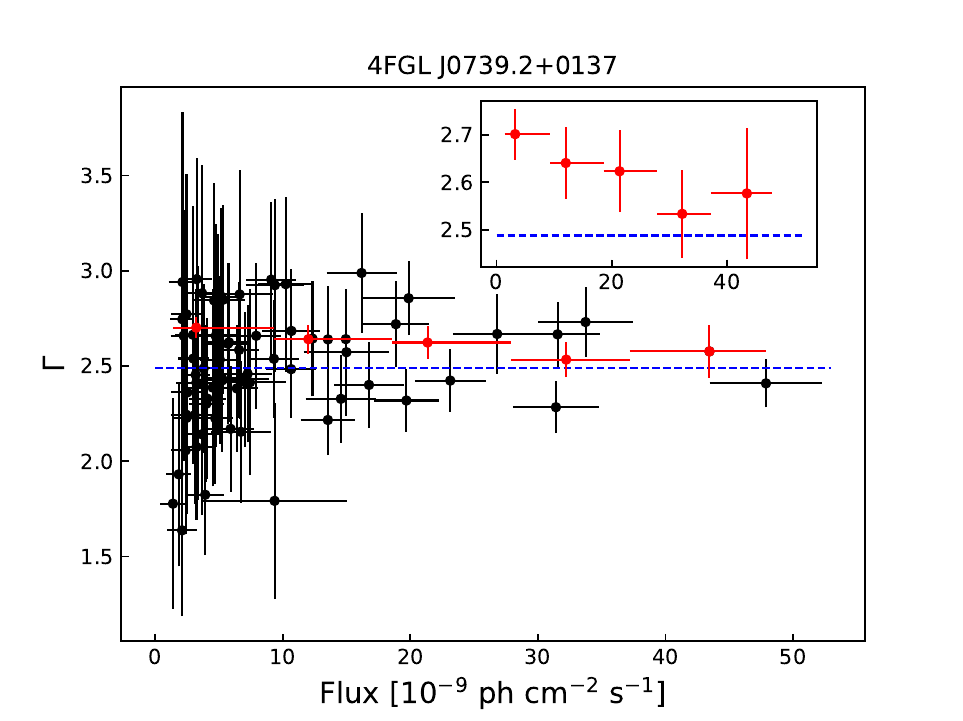}

\includegraphics[width=0.31\textwidth]{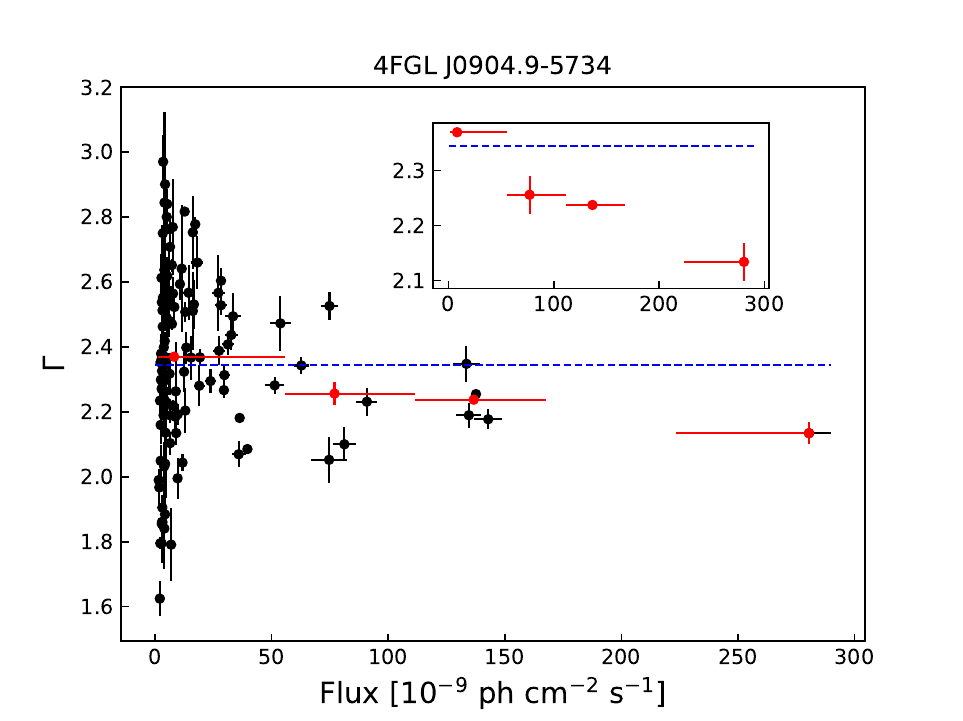}
\includegraphics[width=0.31\textwidth]{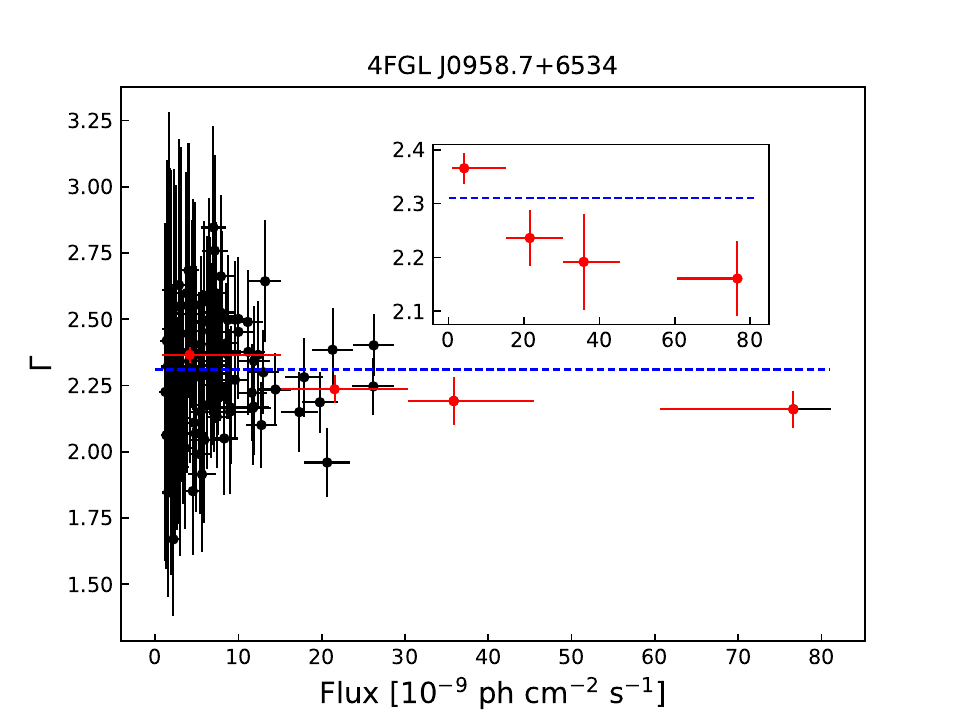}
\includegraphics[width=0.31\textwidth]{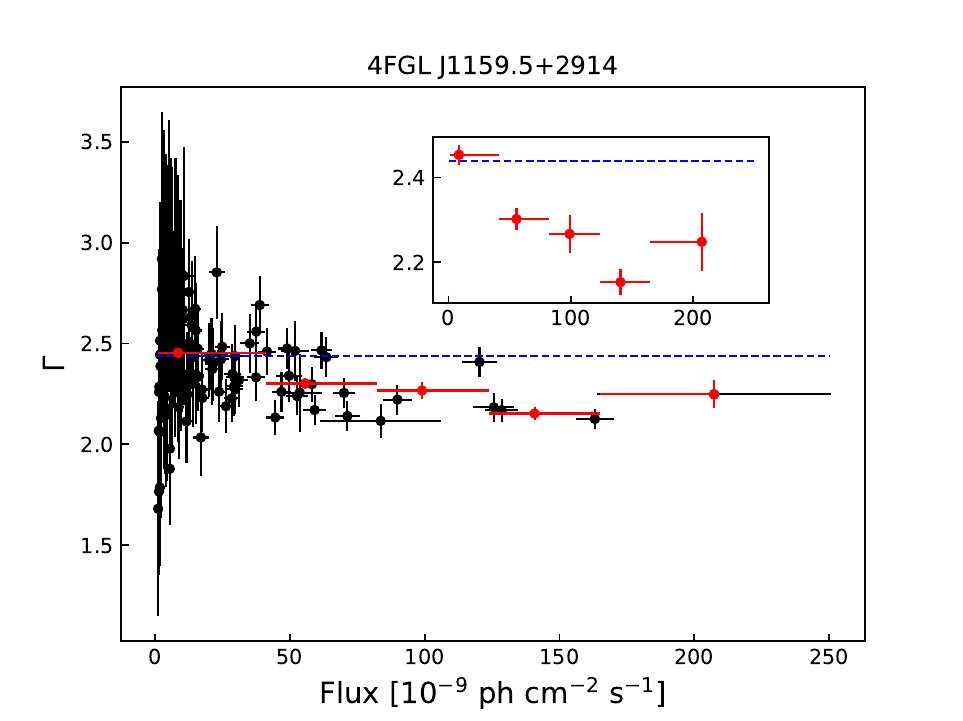}

\includegraphics[width=0.31\textwidth]{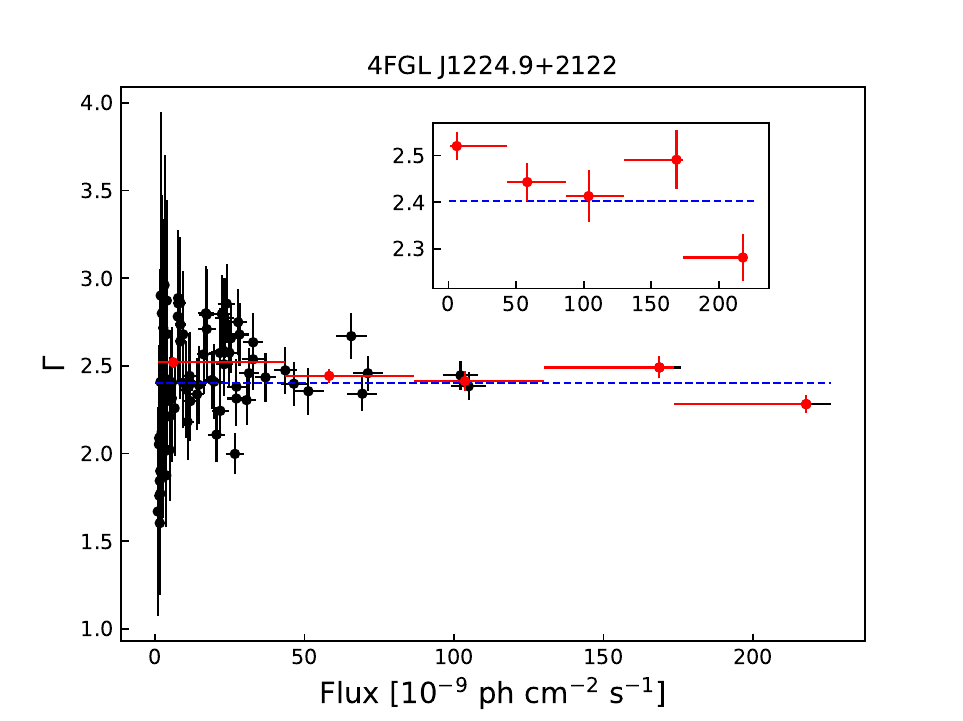}
\includegraphics[width=0.31\textwidth]{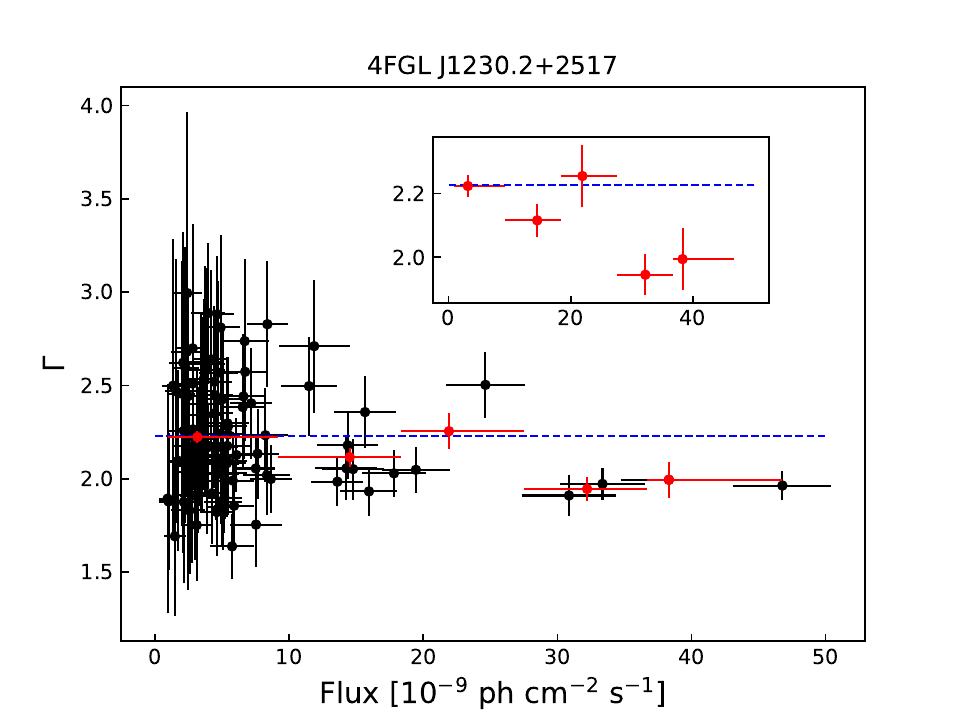}
\includegraphics[width=0.31\textwidth]{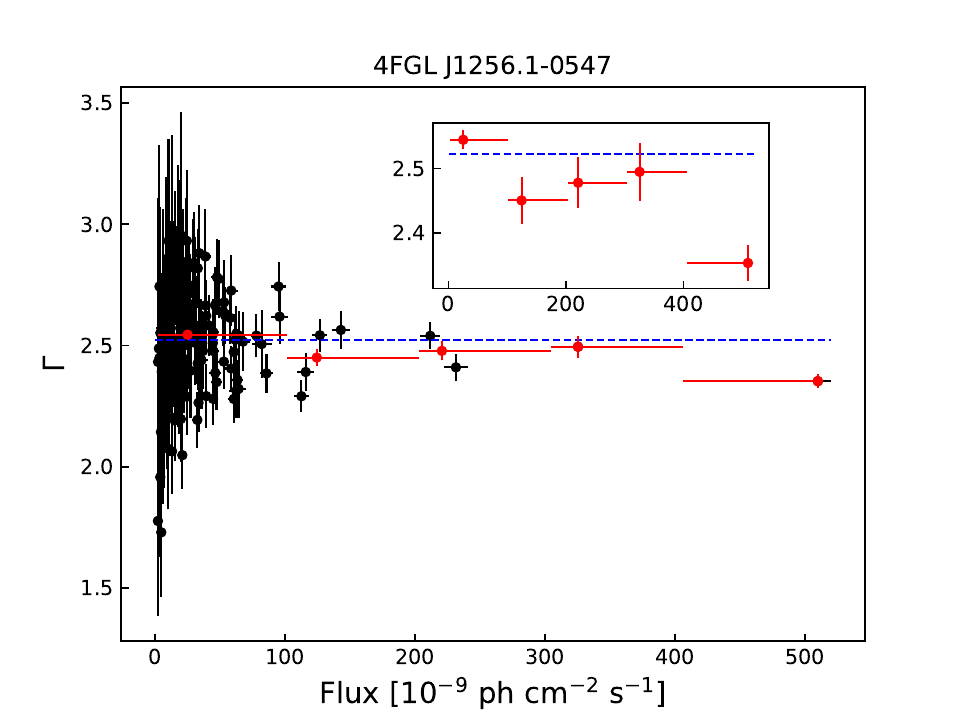}

\includegraphics[width=0.31\textwidth]{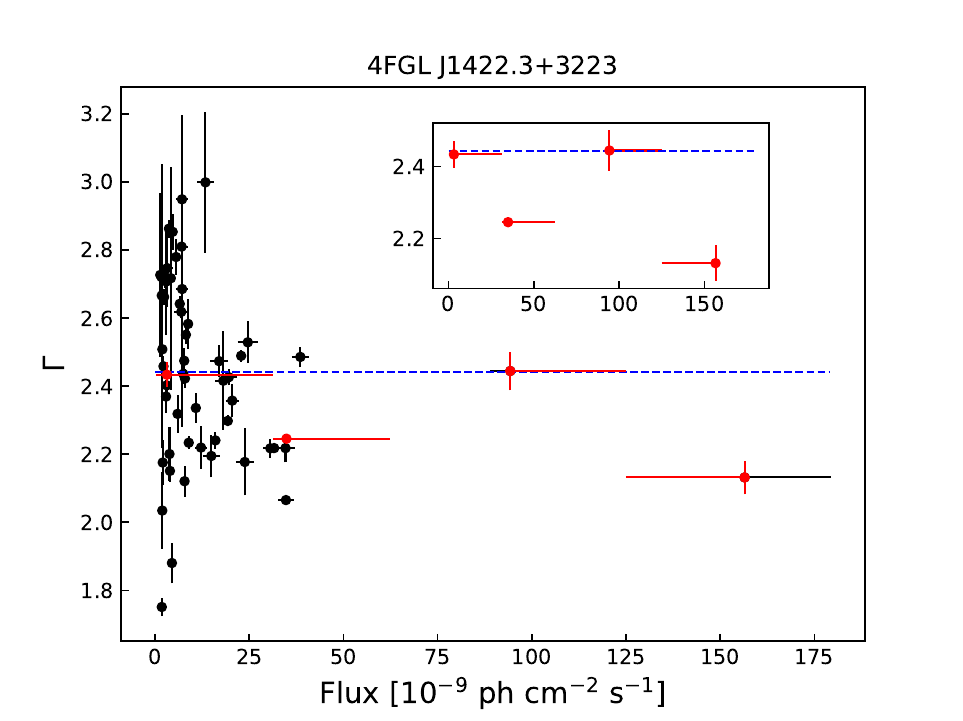}
\includegraphics[width=0.31\textwidth]{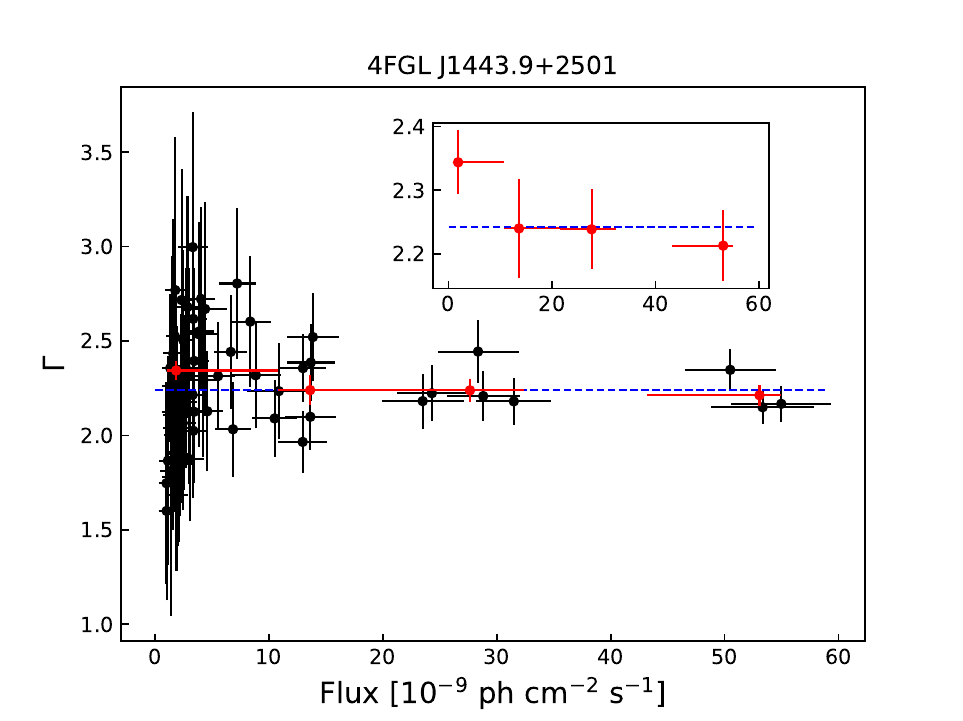}
\includegraphics[width=0.31\textwidth]{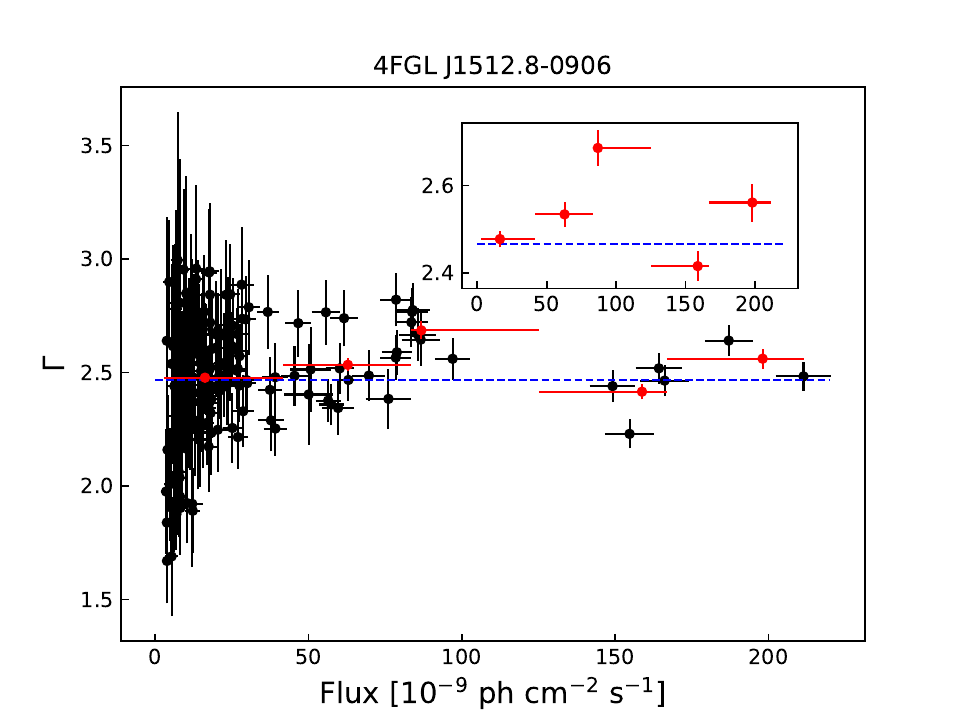}

\includegraphics[width=0.31\textwidth]{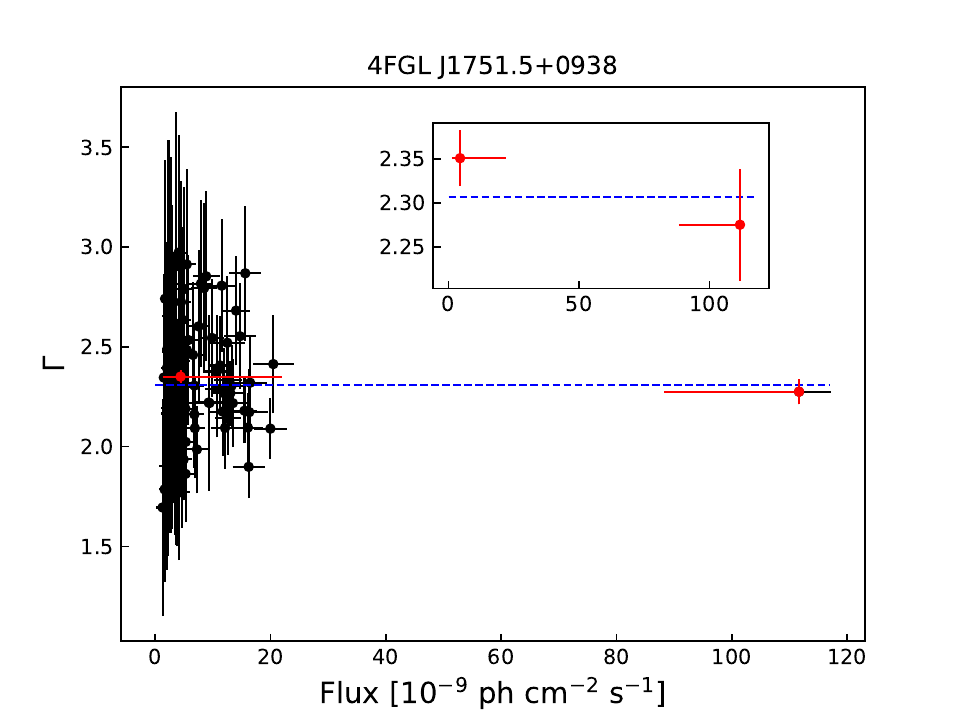}
\includegraphics[width=0.31\textwidth]{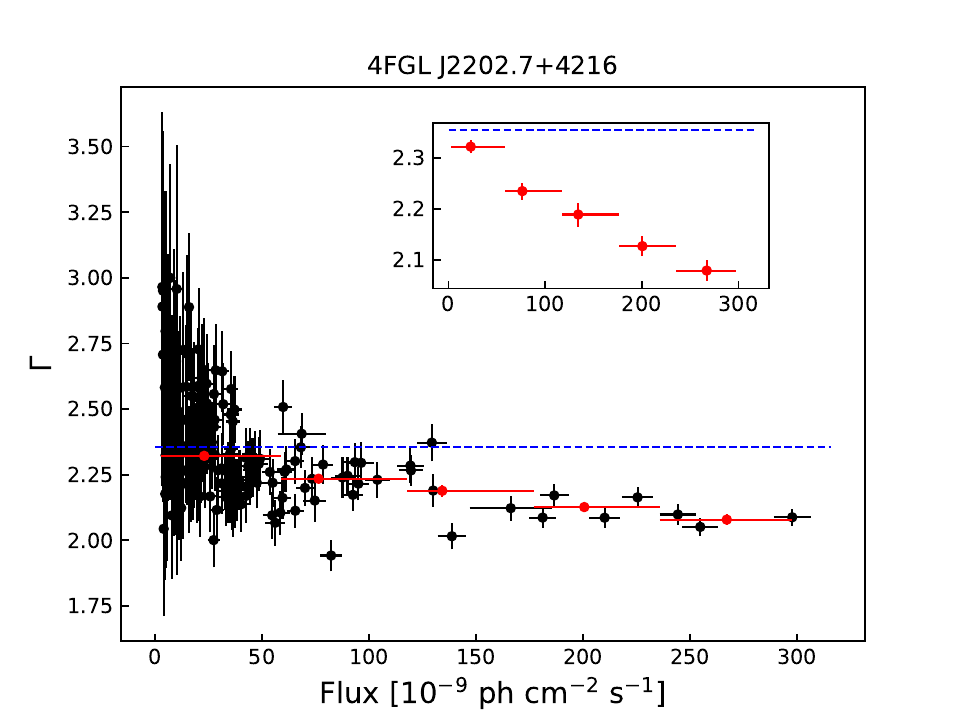}
\caption{1--300 GeV photon indices and fluxes of blazars with bright flares. Only data points with TS $>$ 9 are plotted. The blue dashed horizontal lines indicate the average photon indices of those data. The insets show the photon index resulting from an analysis where photons were sorted in five bins using 5-day fluxes plotted vs. the 5-day flux (red points).}
\label{fig:flux_index}
\end{figure*}

We note that 28 of our TeV blazar samples were reported in the second \textit{Fermi} all-sky variability analysis catalog (2FAV, \citealt{2fav}). The analysis of 2FAV was ran in weekly time bins using the first 7.4 years of \textit{Fermi} data in two independent energy bands, 100--800 MeV and 0.8--300 GeV. 
We have checked these light curves to find the GeV outbursts/flares that meet the criterion, which is that a source shows flare flux more than 10 times larger than its flux in quiescent states and the significance compared
to the quiescent light curves is more than 5$\sigma$ simultaneously.
There are 14 sources that show significant outbursts/flares at the GeV band during the \textit{Fermi} campaign, these sources are listed in Table \ref{tab:flare}. 
All of these significant flares have been reported in 2FAV, except for J1422.3+3223.
The photon indices and fluxes of these 14 blazars with bright flares in the 1--300 GeV band are shown in Figure~\ref{fig:flux_index}, for which only the flux data points with TS $>$ 9 were selected for the plot. 
The insets in Figure~\ref{fig:flux_index} display the photon index resulting from an analysis where photons were sorted in five bins in 5-day flux, plotted versus the 5-day flux.
Fluxes and photon indices during their flaring states are listed in columns (4) and (5) of Table~\ref{tab:flare}, respectively.

\begin{figure}
\centering
\includegraphics[width=0.32\textwidth]{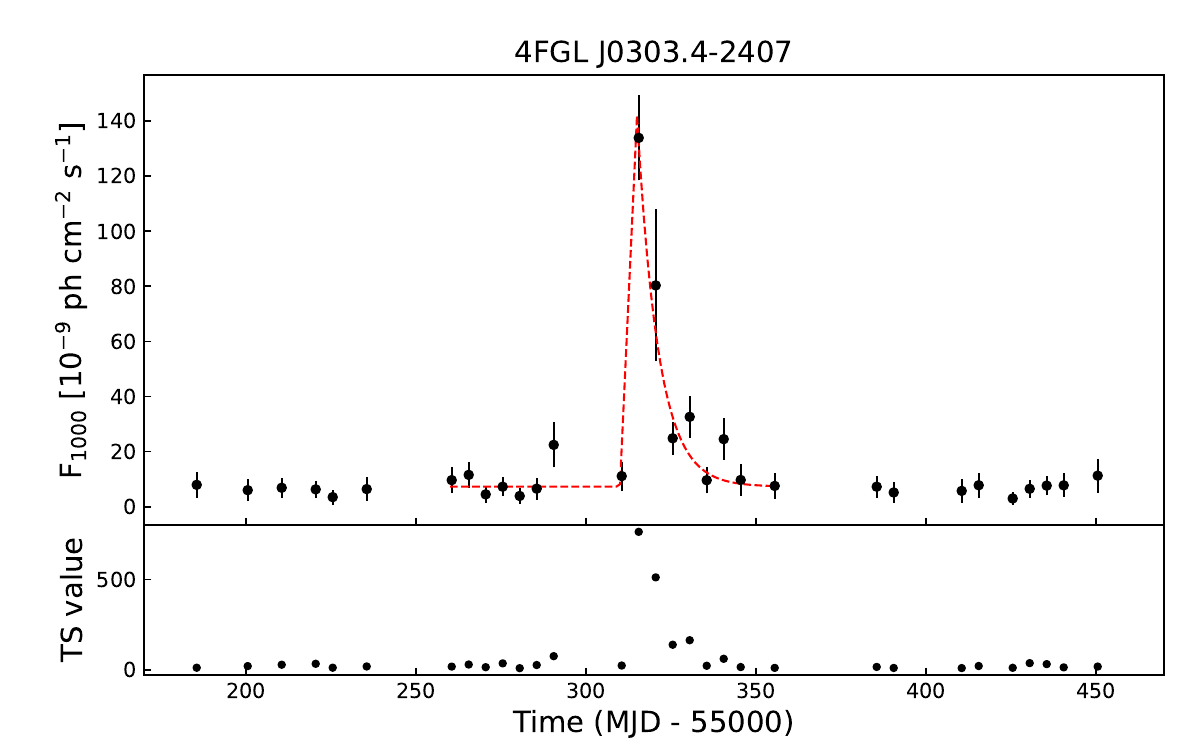}
\includegraphics[width=0.32\textwidth]{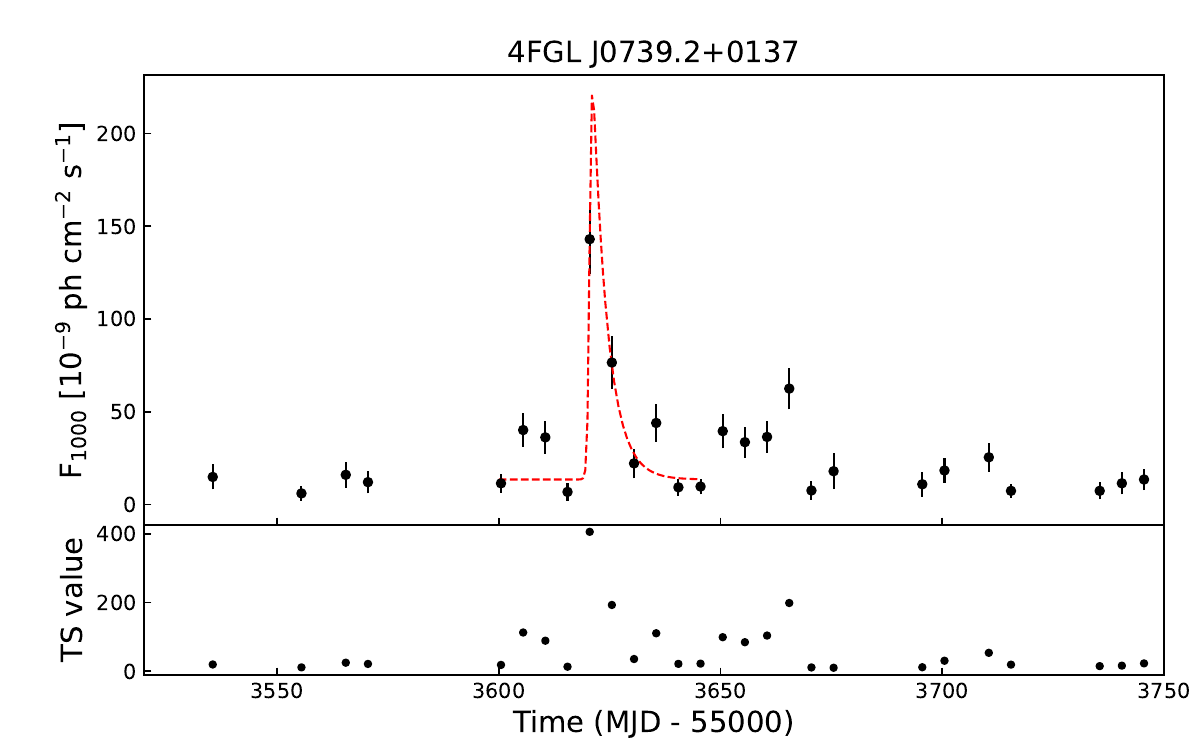}
\includegraphics[width=0.32\textwidth]{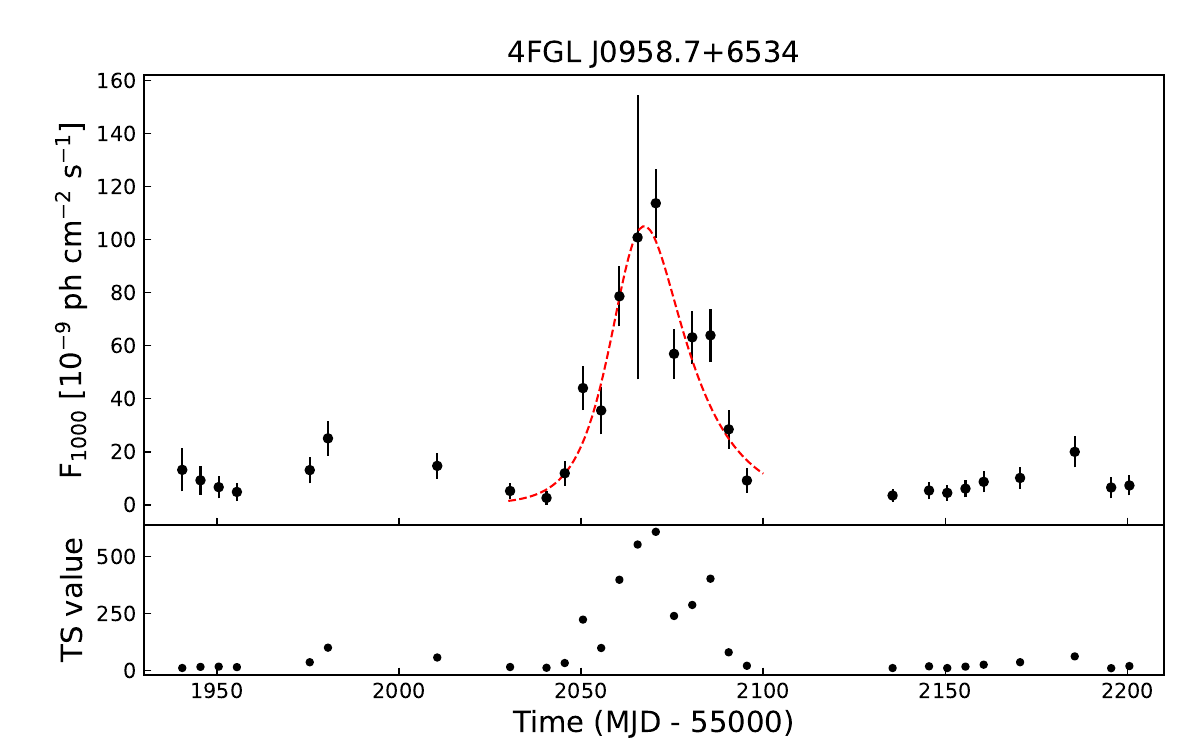}

\includegraphics[width=0.32\textwidth]{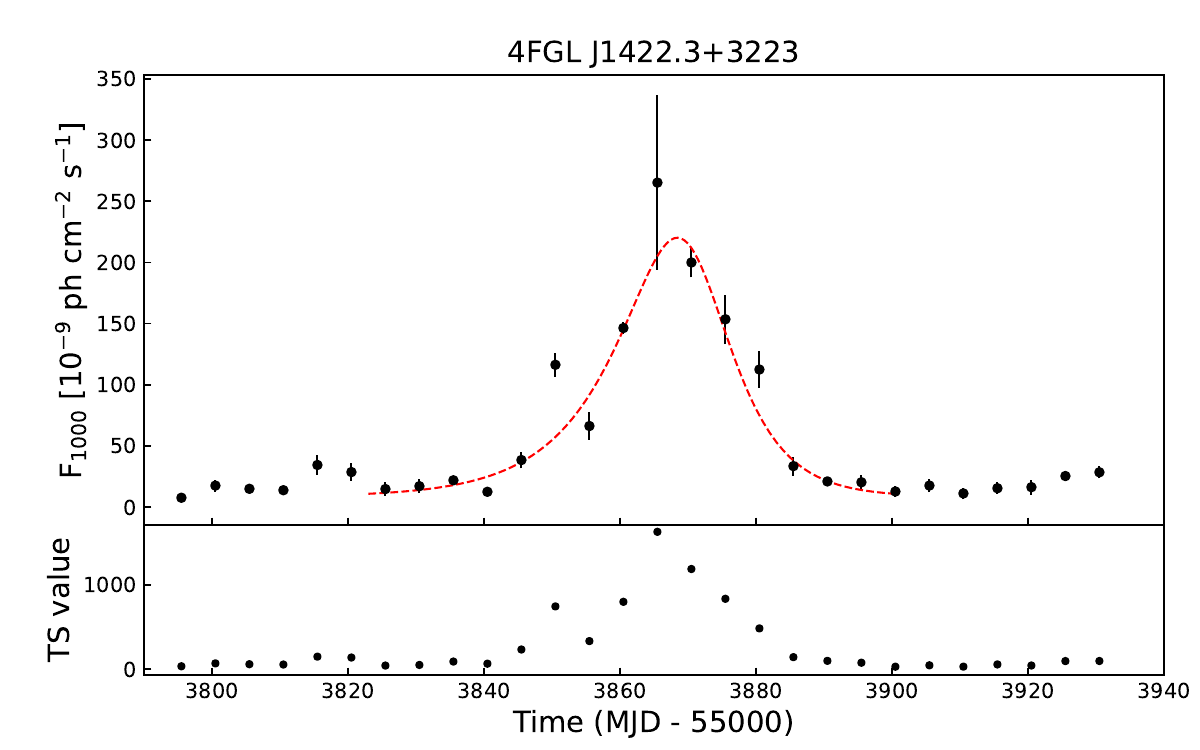}
\includegraphics[width=0.32\textwidth]{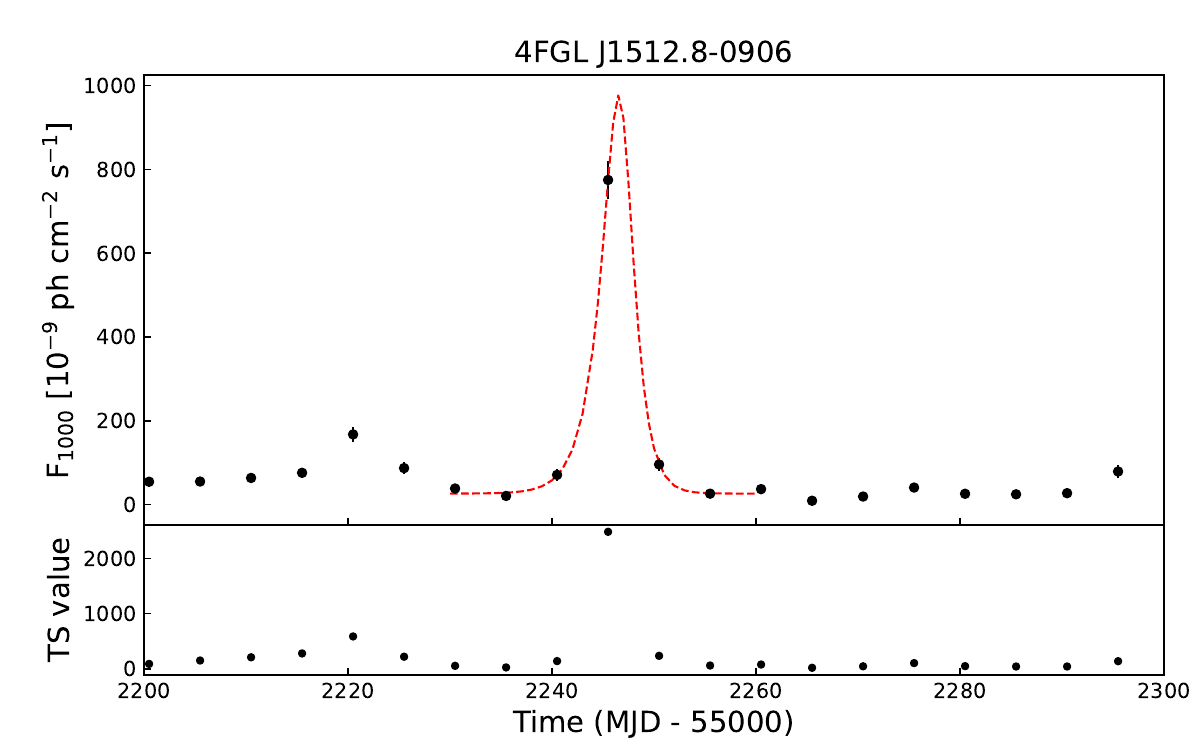}
\includegraphics[width=0.32\textwidth]{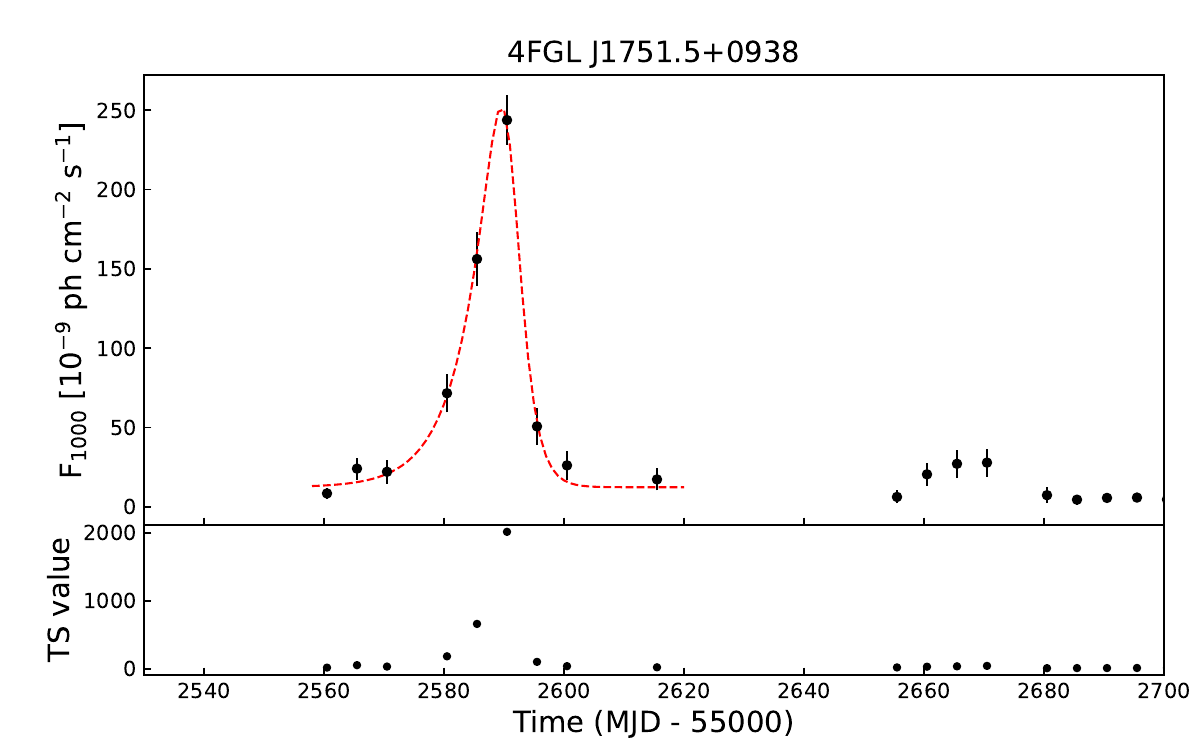}

\caption{Flare profiles for 6 TeV blazars with a single sharp peak in their 5-day binned light curves, an analytic function (dashed red curve) was used to fit the profile.}
\label{fig:profile}
\end{figure}

\begin{figure*}
\centering
\includegraphics[width=0.6\textwidth]{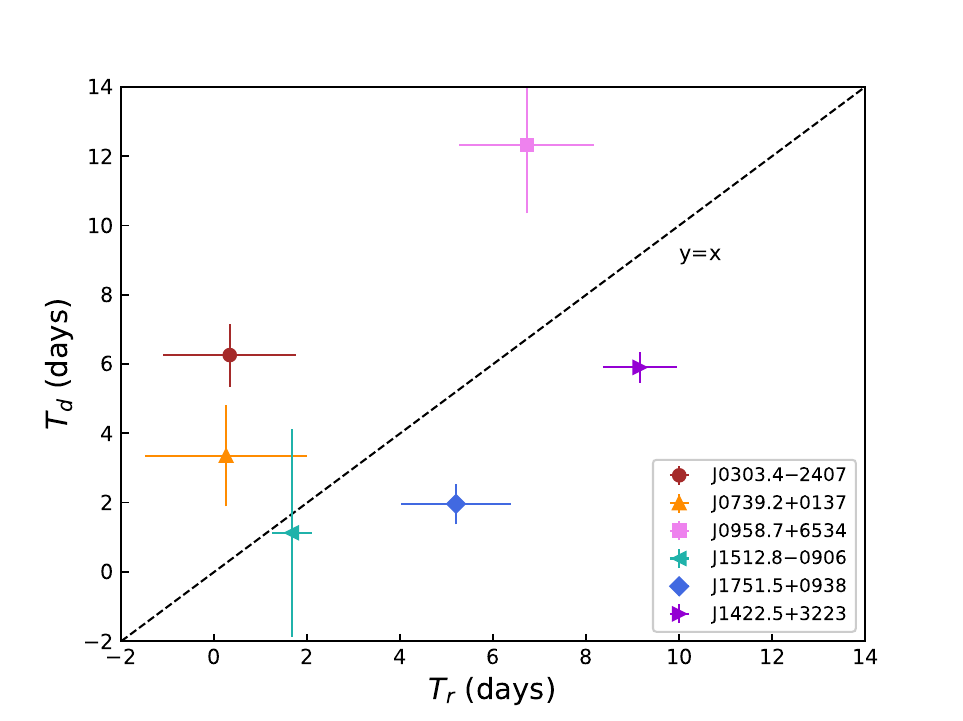}
\caption{Distribution of rise and decay time for the 6 sharp peak flare profiles fitted to the data.}
\label{fig:rise_decay}
\end{figure*}

For these 14 blazars with bright flares, we constructed light curves in the 5-day bin in their flaring states. There are 6 blazars that showed a clear single sharp peak profile contained in the flare that meet the criterion, which is that its flare flux shows more than 12 times larger than its flux in quiescence and the significance compared to the quiescent states is more than 4$\sigma$ simultaneously. 
We also searched intra-day flares and only found 4FGL J1256.1$-$0547 (3C 279) had minute-scale variability in 2018, and this result has been reported in our previous work \citep{wggpasp}. 
We determined the properties of the 6 single sharp peak cases by fitting their profiles with a formula given by
\begin{equation}
    F(t)=F_{\mathrm{c}}+F_0\left(e^{\left(t_0-t\right) / T_{\mathrm{r}}}+e^{\left(t-t_0\right) / T_{\mathrm{d}}}\right)^{-1},
\end{equation}
where $F_{\mathrm{c}}$ and $F_{\mathrm{0}}$ are the constant flux and height of a peak, respectively, $t_{\mathrm{0}}$ is the flux peak time, $T_{\mathrm{r}}$ and $T_{\mathrm{d}}$ are used to measure the rise and decay time in units of day. We show the flare profiles in Figure~\ref{fig:profile} for flares with a single sharp peak, and the distribution of rise and decay time in Figure~\ref{fig:rise_decay}.

We can calculate the flare asymmetry parameter following \citet{chat+12} as 
\begin{equation}
    k=\frac{T_{\mathrm {r}}-T_{\mathrm {d}}}{T_{\mathrm {r}}+T_{\mathrm {d}}},
\end{equation}
the results are listed in column (9) of Table~\ref{tab:flare}, while $k<0$ indicates a fast-rise exponential-decay (FRED) type flare. 
Approximately, $k<-0.3$ indicates faster rise than decay, $k>0.3$ indicates faster decay than rise, while $-0.3<k<0.3$ indicates a symmetric profile, $k=0$ for exactly symmetric flares. 
Among the 6 sharp peak flares, 4FGL J0303.4$-$2407 and 4FGL J0739.2+0137 show FRED behavior, 4FGL J1751.5+0938 shows the opposite, and the other three show symmetric profiles. 
\citet{chat+12} showed the distribution of the flare asymmetry parameter ($k$) for the optical and \gr\ flares with a sample of six blazars, which indicated that most of the flare profiles are symmetric at both wave bands. 
\citet{Abdo2010ApJ722} gave a systematic analysis of a larger sample of 106 objects by using the first 11 months of data of the \textit{Fermi} survey and found only 4 sources with markedly asymmetric flares.

\begin{deluxetable*}{lcccccccc}
\label{tab:flare}
\tabletypesize{\scriptsize} \tablecaption{Fluxes, photon indices, fitting results for the flaring sharp peaks, and asymmetry parameters of the bright GeV flares of 14 TeV blazars}
\tablewidth{0pt}
\tablehead{
\colhead{4FGL name} &
\colhead{$z$} &
\colhead{Class} &
\colhead{$f_{\gamma}$ (${\rm ph} \cdot {\rm cm}^{-2} \cdot {\rm s}^{-1}$)} &
\colhead{$\Gamma$} &
\colhead{$\alpha$} &
\colhead{$T_{\rm r}$ (d)} &
\colhead{$T_{\rm d}$ (d)} & 
\colhead{$k$} \\
\colhead{(1)} &
\colhead{(2)} &
\colhead{(3)} &
\colhead{(4)} &
\colhead{(5)} &
\colhead{(6)} &
\colhead{(7)} &
\colhead{(8)} &
\colhead{(9)} 
} 
\startdata
J0221.1$+$3556	& 0.944	& FSRQ	& 9.51e-08 $\pm$ 3.05e-09 & 2.54 $\pm$ 0.03	&		&		          &                  &           \\
J0303.4$-$2407	& 0.26	& IBL	& 5.15e-08 $\pm$ 3.89e-09 & 1.97 $\pm$ 0.08	&		& 0.33 $\pm$ 1.43 & 6.26 $\pm$ 0.91  & -0.90      \\
J0739.2$+$0137	& 0.189	& FSRQ	& 3.78e-08 $\pm$ 2.99e-09 & 2.55 $\pm$ 0.12	& 4.20	& 0.26 $\pm$ 1.74 & 3.36 $\pm$ 1.47  & -0.86     \\
J0904.9$-$5734	& 0.695	& BCU	& 2.09e-07 $\pm$ 5.22e-09 & 2.19 $\pm$ 0.02	& 2.76  &		          &                  &           \\
J0958.7$+$6534	& 0.367	& IBL	& 7.66e-08 $\pm$ 4.58e-09 & 2.16 $\pm$ 0.07	& 2.64  & 6.72 $\pm$ 1.45 & 12.32 $\pm$ 1.97 & -0.29      \\
J1159.5$+$2914	& 0.729	& FSRQ	& 1.74e-07 $\pm$ 2.38e-08 & 2.20 $\pm$ 0.05	& 2.80	&		          &                  &            \\
J1224.9$+$2122	& 0.435	& FSRQ	& 1.63e-07 $\pm$ 1.18e-08 & 2.39 $\pm$ 0.06	& 3.56	&		          &                  &            \\
J1230.2$+$2517	& 0.135	& IBL	& 3.08e-08 $\pm$ 2.64e-09 & 2.06 $\pm$ 0.09	& 2.24	&		          &                  &           \\
J1256.1$-$0547	& 0.536	& FSRQ	& 3.52e-07 $\pm$ 8.72e-09 & 2.44 $\pm$ 0.04	& 3.76	&		          &                  &           \\
J1422.3$+$3223	& 0.682	& FSRQ	& 1.25e-07 $\pm$ 1.41e-08 & 2.29 $\pm$ 0.05	& 3.16	& 9.15 $\pm$ 0.79 & 5.90 $\pm$ 0.44  & 0.22       \\
J1443.9$+$2501	& 0.939	& FSRQ	& 4.04e-08 $\pm$ 1.95e-09 & 2.23 $\pm$ 0.06	& 2.92	&		          &                  &           \\
J1512.8$-$0906	& 0.36	& FSRQ	& 1.78e-07 $\pm$ 4.90e-09 & 2.49 $\pm$ 0.04	&		& 1.67 $\pm$ 0.43 & 1.13 $\pm$ 2.99  & 0.19      \\
J1751.5$+$0938	& 0.322	& LBL	& 1.12e-07 $\pm$ 5.62e-09 & 2.27 $\pm$ 0.06	& 3.08	& 5.20 $\pm$ 1.18 & 1.96 $\pm$ 0.59  & 0.45      \\
J2202.7$+$4216	& 0.069	& IBL	& 2.34e-07 $\pm$ 4.17e-09 & 2.10 $\pm$ 0.02	& 2.40	&		          &                  &            \\
\enddata 
\tablecomments{(1): 4FGL name; (2): The redshift; (3): The classification that determined based on the synchrotron peak frequency; (4): Fluxes during flaring states; (5): Photon indices during flaring states; (6): The electron spectra index in subsection \ref{subsec:spectral_behavior}; (7): The rise time fitting results for the flaring sharp peaks in units of day; (8): The decay time fitting results for the flaring sharp peaks in units of day; (9): The flare asymmetry parameter.}
\end{deluxetable*}

\section{Discussion} \label{sec:4}
\subsection{The connection with the TeV band}\label{sec:4.3}
We have checked the coincidence between \lat\ GeV detections and TeV detections of the sample. 
There are 22 sources in Table~\ref{tab:sample} that have been detected TeV emission during the flaring states observed by \lat\ , and 56 sources were in low-state. 
J0509.4+0542 (TXS 0506+056) was detected at VHE by MAGIC and VERITAS \citep{ans+18,abe+18,acc+22}. 
It was in an active flaring state around the arrival of the high-energy neutrino IceCube-170922A \citep{IceCube2018Sci_2}. 
While \citet{gar+19} found another blazar GB6 J1040+0617, in spatial coincidence with a neutrino in this sample and the chance probability of 30\% after trial correction, indicating the source of this neutrino remains unknown. 
J1015.0+4926 was detected in a flaring state at VHE by MAGIC during February$-$March 2014\citep{ahn+16}, \lat\ observation was coincident with the TeV detection, and the GeV flux reached a level of 6.5 times higher than its low-state. 
J1058.6+2817, \lat\ and MAGIC successively reported its flaring activity during March$-$April 2021 \citep{ang+21,bla+21}. 
J1217.9+3007, its multiwavelength observations with VERITAS and \lat\ showed a well-connected high flux state in February 2014 \citep{abe+17}. 
J1728.3+5013, \citet{arch+15} reported the first detection of \gr\ flaring activity at VHE from this blazar, the flaring flux is about five times higher than its low-state. 
\lat\ detected this source with mild flare and it was observed a photon of energy more than 300 GeV as reported in \citet{magic2020}. 
VERITAS detected VHE emission from J1813.5+3144 with the similar flux reported by MAGIC \citep{ben+22}, during the active state observed by \lat\ in 2020.
J2000.0+6508 was reported to show flaring activity during June–July 2016 by \lat\ and MAGIC \citep{magic2000+}. 
J2001.2+4353 showed a significant TeV detection on 2010 July 16 reported in \citet{ale+14}, during the flaring activity observed by \lat\ . 
J2243.9+2021 was also active in high-energy (HE) during the time of VHE detection with the flux larger than the four-year-averaged flux reported in 3FGL \citep{abey+17}. 

As for the 14 blazars that have bright flares, their flaring LAT states are all coincident with the TeV detections, except for J2202.7+4216. 
The detailed results are as follows:
\begin{itemize}
\item[ ] J0221.1+3556, which was detected by MAGIC in July 2014 was in its minor-flare state in \lat\ observations, while its major-flare state was in September 2012. 
The TeV detection was during the expected delayed component of the \lat\ flare \citep{ahn+16}. 
\item[ ] J0303.4$-$2407 was detected in a high-state defined lasting from MJD 55312 (April 26, 2010) to MJD 55323 (May 5, 2010) reported by \citet{hess+13}, which is also coincident with our result. 
However, the flaring TeV state in November 2011 is during the low-state observed by \lat\ . 
\item[ ] J0739.2+0137, its H.E.S.S. observation was triggered on the basis of the detection of a \lat\ flare, resulting in the detection of VHE \gr\ emission during the night of February 19, 2015. 
Therefore its flaring TeV state was coincident with flaring \lat\ state \citep{hess+2020}. 
\item[ ] J0904.9$-$5734, H.E.S.S. observed a significant detection of VHE emission on 2020 April 13 \citep{wagner2020}, which is during the flaring state MJD 58931-58970 observed by \lat\ . 
\item[ ] J0958.7+6534 was detected VHE \gr\ emission by \citet{magic18aa} during the time period (2015 February, 13/14, or MJD 57067). 
While \lat\ detected a 51 GeV photon from a very close position (0.013$^\circ$) of J0958.7+6534 on MJD 57066.98, indicating the coincidence with the MAGIC VHE detection \citep{tan+16}. 
\item[ ] J1159.5+2914, its time evolution of flux detected by \lat\ was similar to VHE lightcurve in its flaring states of 2017 and 2021 \citep{hir+18,ada+22}. 
\item[ ] J1224.9+2122, MAGIC detected its VHE emission around MJD 55364.9 (June 2010), this coincided with the flaring state at GeV energies \citep{hayes11}. 
\item[ ] J1230.2+2517, \citet{ach+23} reported the follow-up multiwavelength observations to the discovery of VHE emission with VERITAS, showing the flaring states in HE and VHE are coincident. 
\item[ ] J1256.1$-$0547 (3C 279) has been extensively studied for its variability properties. 
For its GeV flare in 2015, the H.E.S.S. observation led to a clear detection during the end of the \lat\ flaring state \citep{hess2019aa,pit+18}. 
For its GeV flare in 2018, intense VHE flares were observed over multiple days after the end of the HE flares \citep{eme+19}. 
\item[ ] J1422.3+3223 was detected by MAGIC during its high-state observed by \lat\ , indicating the coincidence between the TeV detection and the flaring LAT state \citep{magic2021aa}.
\item[ ] J1443.9+2501 was reported the VHE detection in the flaring state observed by \lat\ \citep{abey+15,ahnen2015}.
\item[ ] J1751.5+0938, H.E.S.S. detected a flaring flux increase in 2016 \citep{sch+17}, which is consistent with the flaring state observed by \textit{Fermi}-LAT.
\item[ ] J1512.8$-$0906, the AGILE results of its flare in 2009 reported in \citet{dam+11} are in agreement with the \lat\ results presented in \citet{2010ApJ1425A}. For the subsequent flares in 2012 and 2015, the HE \gr\ light curve showed a mild flux variation compared to the strong flare at VHE energies \citep{zac+19,hess2021}. 
\end{itemize}

\subsection{Variability analysis}
\subsubsection{Fractional variability and flare profile}
\begin{figure}
\centering
\includegraphics[width=0.6\textwidth]{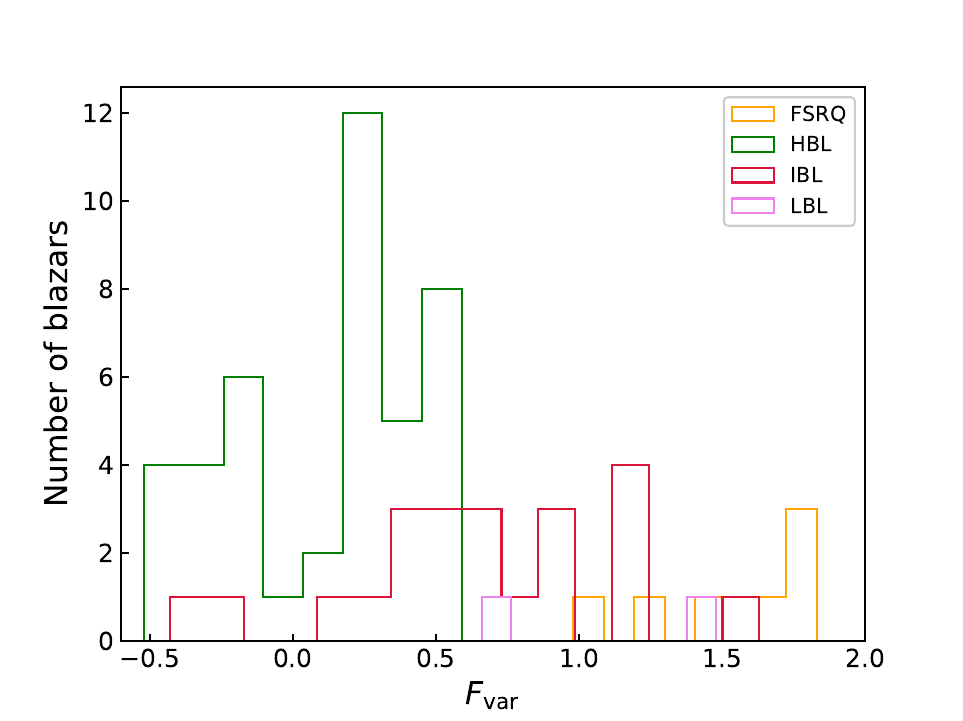}
\caption{Distribution of the fractional variability $F_{\mathrm{var}}$ for the light curves of FSRQs, HBLs, IBLs, and LBLs.}
\label{fig:fvar}
\end{figure}

Variability is one of the main characteristics of blazar that has been studied in multi-bands \citep{Urry1996ASPC, Dermer1999APh, Fan1999MNRAS, Singh2020AN, Webb2021Galax, Yuan2022ApJS262, Otero2022MNRAS}.
\citet{Abdo2010ApJ722} suggested more than 50\% \textit{Fermi} detected bright blazars are found to be variable with high significance, FSRQs and LBLs show higher variation amplitudes than the other blazars.
We quantified the variability utilizing the fractional variability parameter $F_{\mathrm{var}}$, $F_{\mathrm{var}}$ can be described as \citep{vau03}
\begin{equation}
F_{\mathrm{var}}=\sqrt{\frac{S^2-\left\langle\sigma_{\mathrm{err}}^2\right\rangle}{\left\langle F_\gamma\right\rangle^2}},
\end{equation}
where $S^2$ is the variance of the flux, $\left\langle\sigma_{\mathrm{err}}^2\right\rangle$ is the mean square value of uncertainties, $\left\langle F_\gamma\right\rangle$ is the mean photon flux. 
Negative values of $F_{\mathrm{var}}$ indicate very small or absent variability and/or slightly overestimated errors. We derived the mean values of $F_{\mathrm{var}}$ are 1.54 $\pm$ 0.02, 0.12 $\pm$ 0.15, 0.65 $\pm$ 0.06, 1.07 $\pm$ 0.04 for the FSRQs, HBLs, IBLs, and LBLs, respectively.  
The resulting values indicate that the flux of the FSRQs showed significantly stronger variability than that of the BL Lacs. 
As the synchrotron peak frequency decreases, the $F_{\mathrm{var}}$ value generally becomes larger. 
Here we presented a histogram of $F_{\mathrm{var}}$ values for the FSRQs, HBLs, IBLs, and LBLs in Figure~\ref{fig:fvar}. 
\citet{bhatta2020} presented an analysis of a sample of 20 powerful blazars (12 BL Lacs and 8 FSRQs) with 10 yr \lat\ data, they obtained that the mean $F_{\mathrm{var}}$ value of BL Lacs is 0.58 and that of the FSRQs is 0.96. 
The results show that in general FSRQs are more variable than BL Lac sources in their sample, which is compatible with ours. 
Similar future studies involving larger samples should be carried out for a stronger conclusion. 
For the individual source, our result of S5 0716+714 is consistent with that reported in \citet{bhatta2016}, the $F_{\mathrm{var}}$ values are 0.65, 0.57, 0.58, 0.53 for BVRI filters versus our 0.59.

Besides, we found 14 TeV blazars (8 FSRQs, 1 LBL, 4 IBLs, and 1 BCU) with outbursts/flares, 6 out of the 14 flares show sharp peak profiles in flares.
Based on the sharp peak profiles, we notice 4FGL J0303.4$-$2407 and 4FGL J0739.2+0137 show a fast-rise-slow-decay subflare.
This asymmetry can be related to the particle acceleration mechanism in the jet, a fast rise could result from an effective particle acceleration at the shock front and slow decay may be interpreted as the weakening of the shock \citep{Sokolov2004ApJ613, Tolamatti2022APh13902687} or from the injection of energetic particles on a shorter timescale than the cooling process timescales \citep{Acharyya2021}. 
While 4FGL J1751.5+0938 shows a slow-rise-fast-decay subflare, which may be associated with an efficient cooling process. 

\subsubsection{Flux distributions}
\begin{figure}
\centering
\includegraphics[width=0.32\textwidth]{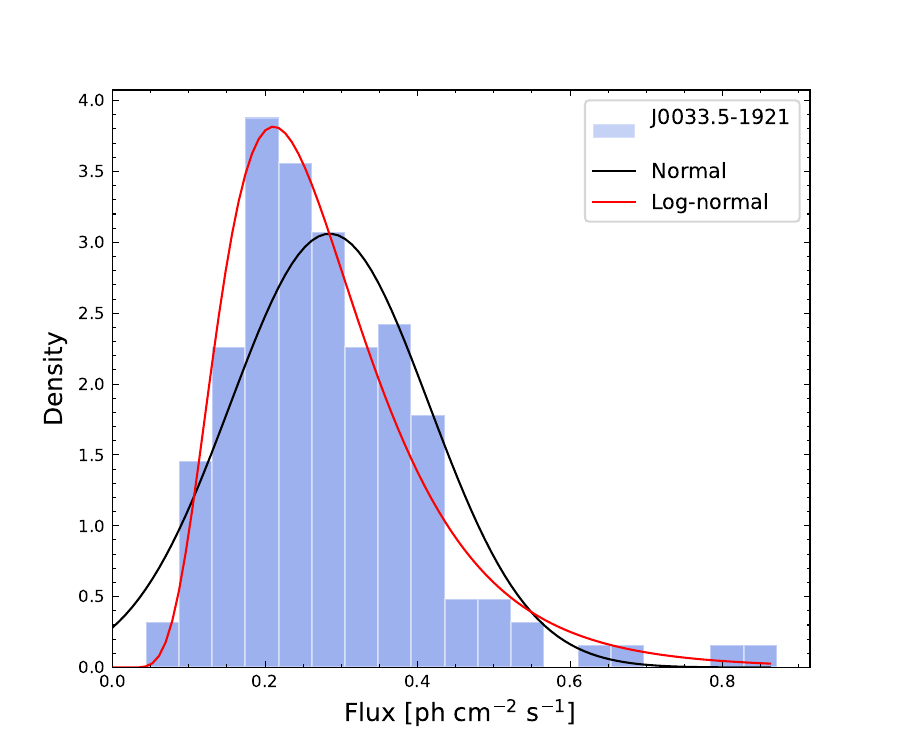}
\includegraphics[width=0.32\textwidth]{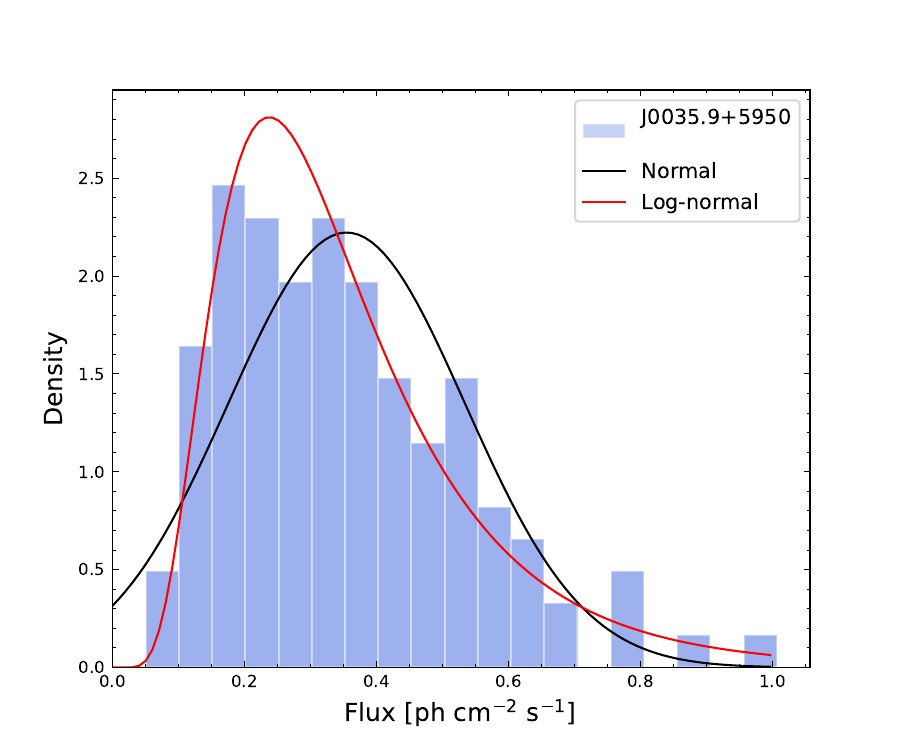}
\includegraphics[width=0.32\textwidth]{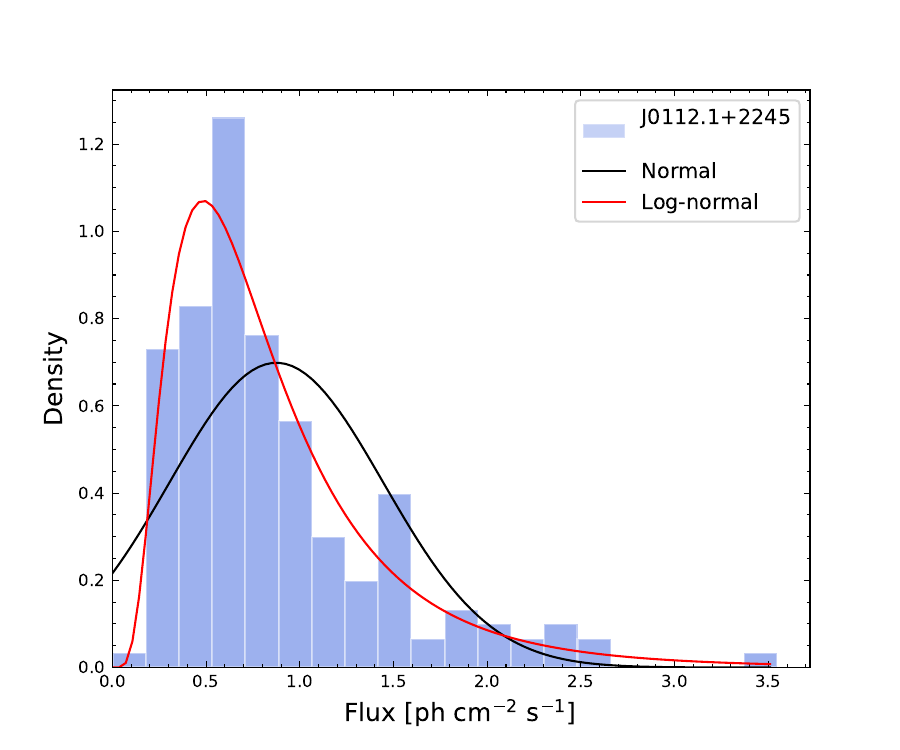}

\caption{Flux distribution of bright blazars in our sample in the GeV
    band. The black and red curves correspond to normal and log-normal
    fits respectively. Only three items are presented here. The complete 
    figure set (41 images) is available in the online journal.}
\label{fig:lognormal}
\end{figure}

The analysis of flux distribution helps us to determine whether the variability is caused by multiplicative or additive mechanisms. Evidence for log-normality in blazars in \gr\ on different timescales has been reported for different sources (e.g., \citealt{kush+17,sinha+17,bhatta2020}). 
Similarly, the log-normal flux distribution of blazars was seen in 3LAC \citep{3lac+15}). 
\citet{shah+18} studied the flux distribution features of the selected 38 brightest \textit{Fermi} blazars using the data collected during more than 8 years and found that the flux distribution for 35 blazars supports a log-normal distribution, implying a multiplicative perturbation linked with the emission process. 
Using a large sample of 1414 variable blazars from the \lat\ LCR catalog, \citet{wangna2023} thoroughly investigated the \gr\ flux distribution and statistical properties, and compared the flux distributions with normal and log-normal distributions. 
Their results showed that the probability of not rejecting log-normal is 42.05\%. 
We constructed histograms of the observed LAT GeV flux and fitted them to two different probability density functions (PDFs), a normal distribution and a log-normal distribution, and compared the results of $\chi^2$. 
To ensure sufficient data points for fitting the flux distribution, we selected the 41 bright blazars mentioned in the subsection~\ref{sec:3.1}. 
According to the chi-squared values from the fit, our results show that all of the bright blazars support a log-normal distribution rather than a normal distribution, which is also consistent with the results of previous studies. 
As the consistency between TeV detection and LAT observation that discussed in the subsection~\ref{sec:4.3}, the TeV detections correspond to the outlier periods of the flux distribution. 
The parameters of the considered two distributions fitting results and the source flux histograms are shown in Table~\ref{tab:lognormal} and Figure~\ref{fig:lognormal}, only several items are presented here. 

\begin{table}[H]
\centering
\fontsize{8}{9}\selectfont
\caption{Parameters of normal and log-normal distribution fitting for the $\gamma$-ray flux distribution of the \lat\ sources. Here $\beta_{\text {slope }}$ gives the slope index result of the periodograms.}\label{tab:lognormal}
\setlength{\tabcolsep}{0.2cm}
\begin{tabular}{ccccccccccc}
\hline\hline
\multirow{2}*{Name} &\multicolumn{3}{c}{{Normal Fit}}& {} &\multicolumn{3}{c}{{Log-normal Fit}} & \multirow{2}*{$\beta_{\text {slope }}$} \\
\cline{2-4}\cline{6-8}
 & Mean& $\sigma$ & $\chi^2$ &  & Mean & $\sigma$  & $\chi^2$ \\
\hline
J0033.5$-$1921 & 0.28 & 0.13 & 2.19 & & -1.35 & 0.45 & 1.12 & 0.27 $\pm$ 0.03 \\
J0035.9+5950   & 0.35 & 0.18 & 2.05 & & -1.17 & 0.52 & 1.16 & 1.34 $\pm$ 0.03 \\
J0112.1+2245   & 0.87 & 0.57 & 1.07 & & -0.33 & 0.63 & 0.31 & 0.52 $\pm$ 0.02 \\
J0136.5+3906   & 0.44 & 0.16 & 1.35 & & -0.87 & 0.35 & 0.42 & 0.58 $\pm$ 0.02 \\
J0221.1+3556   & 0.66 & 1.11 & 0.93 & & -0.80 & 0.71 & 0.09 & 0.57 $\pm$ 0.02 \\
J0222.6+4302   & 1.19 & 0.79 & 0.48 & & -0.01 & 0.58 & 0.16 & 0.91 $\pm$ 0.02 \\
J0303.4$-$2407 & 0.48 & 0.45 & 0.68 & & -0.93 & 0.57 & 0.09 & 0.25 $\pm$ 0.02 \\
\hline\hline
\end{tabular}\\[0.1cm]
\begin{minipage}[]{200mm}
{Table \ref{tab:lognormal} is published in its entirety in the machine-readable format. A portion is shown here for guidance regarding its form and content.}
\end{minipage}
\end{table}

\subsubsection{Flare duty cycle}
The flaring state lasts only a fraction of the observation. 
Here we define the flaring state when any of the light curve's flux points exceeds a certain threshold following the method in \citet{yoshida2023}, $f_\gamma^{\mathrm{th}}$, which is given by
\begin{equation}
    f_\gamma^{\mathrm{th}}=f_\gamma^q+s\left\langle f_\gamma^{\mathrm{err}}\right\rangle,
\end{equation}
where $f_\gamma^q$ is the quiescent level of \gr\ fluxes, $\left\langle f_\gamma^{\mathrm{err}}\right\rangle$ is the average uncertainty of the \gr\ fluxes, and $s$ denotes the significance above the quiescent level in standard deviation units of $\sigma$.
Here we use $s=6$ in this work, and the flaring threshold levels are plotted with dashed grey lines in Figure \ref{fig:30dlc}.
From the light curves, we calculated the flare duty cycle (i.e.,
fraction of time spent in flaring states) for each flare. 
The flare duty cycle, is defined as 
\begin{equation}
    f_{\mathrm{fl}}=\frac{1}{T_{\mathrm{tot}}} \int_{f_\gamma^{\mathrm{th}}} d f_\gamma \frac{d T}{d f_\gamma},
\end{equation}
where $T_{\mathrm{tot}}$ is the total observation time, $f_\gamma$ is the \gr\ photon flux, and $T$ is the time spent at the respective flux level. 
We find that our duty cycle results of the monthly-binned light curves show values ranging from 0.0 to 0.36, and there is no evidence to show that the duty cycle is related to the TeV detection.
Based on monthly-binned light curves of the 2-year \lat\ point source (2FGL) catalog, \citet{2lac} showed that bright blazars have flare duty cycles of about 0.05$-$0.10.  
According to Table 2 in \citet{2fav}, the number of weekly-binned flares detected for each source using the first 387 weeks of \textit{Fermi} observations were presented, and the flare duty cycles appeared to suppress less than $\sim 0.2$. 
\citet{yoshida2023} analyzed 145 gamma-ray bright blazars among the
4FGL catalog, their results showed much broader distributions of flare duty cycles from the weekly-binned light curves, ranging from 0.0 to 0.6. 
Our results of flare duty cycle values are similar with previous studies. 
Due to the vast majority of our results being in the range of 0.0 to 0.2, except for three sources with higher duty cycles (0.36 for J0721.9+7120, 0.26 for J1104.4+3812, and 0.26 for J2202.7+4216.)

\subsubsection{Power spectral densities}
\begin{figure}[H]
\centering
\includegraphics[width=0.32\textwidth]{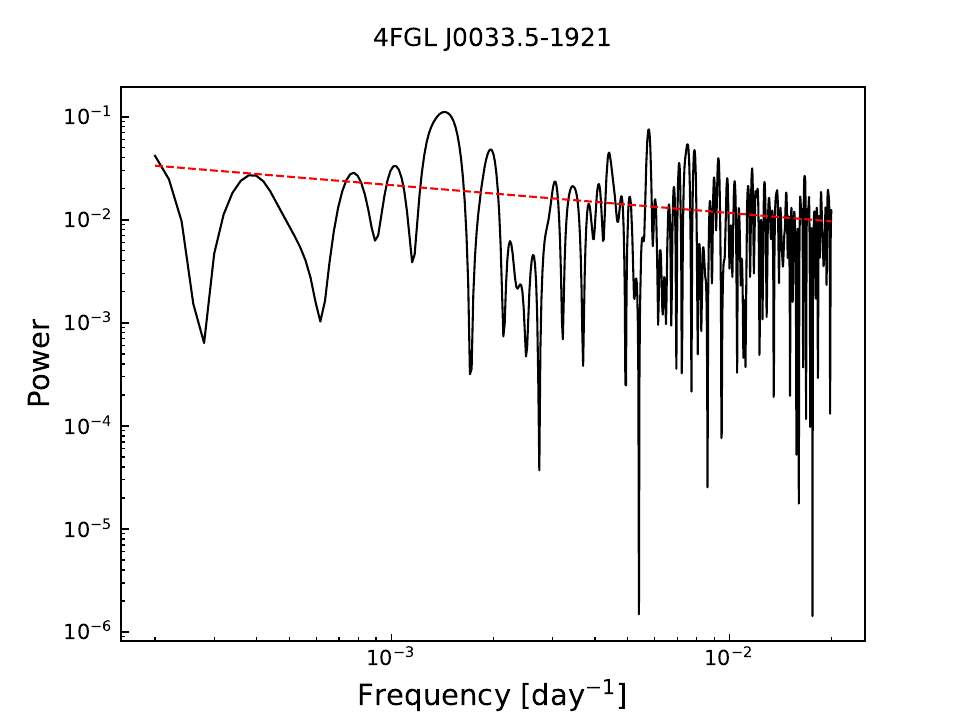}
\includegraphics[width=0.32\textwidth]{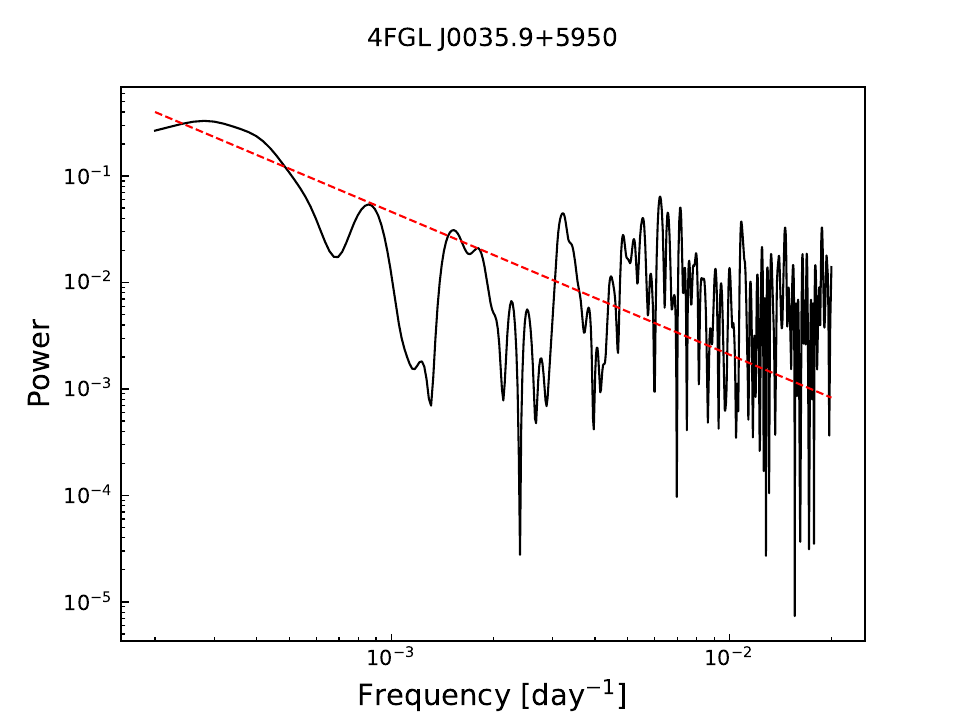}
\includegraphics[width=0.32\textwidth]{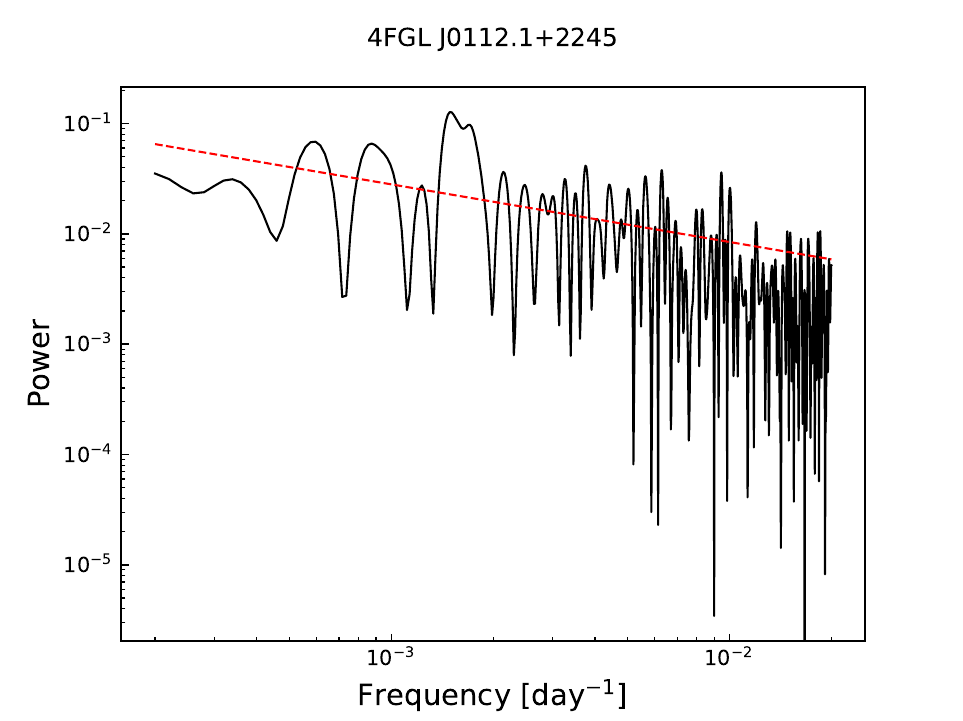}

\caption{PSD fits with power-law for LSPs of the bright blazars. The
    black curve is the raw LSP, and the red dashed line is the best
    fit. Only three items are presented here. The complete figure set 
    (41 images) is available in the online journal.}
\label{fig:psd}
\end{figure}

Power spectral density (PSD) is a mathematical function that characterizes the shape of a source periodogram.
Similarly, in order to ensure the quality of the analysis, we analyzed the periodograms of the monthly binned \gr\ light curves of the 41 bright blazars applying the Lomb-Scargle periodogram (LSP; \citealt{lomb1976,scar+82}). 
For frequency selection of the LSP analysis, the lower limit for the sampled frequencies, which corresponds to the length of the time series is $f_{\min }=1 /\left(t_{\max }-t_{\min }\right)$. 
\citet{eye+99} proposed a meaningful method to assess the Nyquist frequency that would be the upper limit of frequency, $f_{\max}$. 
The approach for selecting the frequency grid is to make each peak in the periodogram to be sampled $n_0$ = 5$-$10 times \citep{van+18}. 
Then the total number of sampling frequencies would be $N=n_0 \frac{f_{\max }}{f_{\min }}$ , and here we employ $n_0$ = 10. 
It is found that the periodograms are consistent with a power-law form of $P(\nu) \propto \nu^{-\beta}$ with the slope index (spectral power index $\beta$) ranging between 0.22$-$1.98. 
The mean PSD slope index of the sources is 0.74 with a standard deviation of 0.41. 
We listed the slope index results in Table \ref{tab:lognormal}, and the plots of the PSD were displayed in Figure \ref{fig:psd}.

\citet{Abdo2010ApJ722} conducted an analysis of the first 11 months of the LAT Bright AGN sample (LBAS), and revealed that the average $\beta$ values of the brightest 22 FSRQs and of the 6 brightest BL Lacs is 1.5 and 1.7, respectively. 
While \citet{ack+11} using 24 months of data and found the $\beta$ value is $\sim1.15 \pm 0.10$, which is somewhat flatter than the results deduced from the LBAS sample. 
\citet{tar+20} presented a comprehensive analysis of the \lat\ 10 yr long light curve modeling of 11 selected blazars by employing various methods. 
They found that the power-law slope index $\beta$ calculated from the Fourier and LSP modeling falls in the range $1 \lesssim \beta \lesssim 2$ mostly. 
Our results of PKS 1510$-$089, PKS 2155$-$304, and Mrk 421 are consistent with \citet{sobo14}. 
They analyzed the \gr\ variability of 13 blazars with a linear superposition of OU processes, for which they found slopes mostly to be $\beta \lesssim 1$. 
\citet{pro+17} obtained $\beta$ = 0.67 for PKS 2155$-$304, while we obtain $\beta$ = 0.65 $\pm$ 0.03. 
Also, our result of 3C 279 is similar to the PSD slopes found by \citet{meyer19}. 
\citet{chat+12} mentioned the average slope of the PSD in R-band of 6 blazars is similar to that found by the \textit{Fermi} team, our result was in agreement for PKS 1510$-$089, but they obtained clearly steeper power-law fits than we did (2.3, 2.2 for 3C 279 and PKS 2155$-$304 versus our 0.75, 0.65).
Compared to these recent results of selecting the several brightest sources, our PSD result at the GeV band is slightly flatter and has a larger range. 
The discrepancies can be caused by the difference in the analysis methods, different binning schemes, sampling interval, and total observation duration of the analyzed light curves or methods of their generation between the works.

\subsubsection{The periodic behaviors}
The periodogram of the light curves can be characterized by a single power-law PSD. 
However, if we closely observe the structures of the periodogram, we may occasionally find peaks at certain frequencies indicating the possible presence of (quasi-) periodic signals in the observations. 
The periodic oscillation in \gr\ band of blazar PG 1553+113 was reported by \citet{Ackermann2015ApJL813}, this source is also contained in our sample and its light curve at the GeV band shows clear periodicity, and has been explained in mechanisms invokes a supermassive binary black hole system \citep{Cavaliere2017ApJ836, Sobacchi2017MNRAS465}. 
Several studies have systematically searched \gr\ QPOs based on 3FGL (e.g., \citealt{pro+17,penil2020}). 
\citet{Penil2022arXiv} made a search for periodicity in a sample of 24 blazars by using 12 well-established methods applied to \textit{Fermi} 12-year data, and found six out of the 24 sources show light curve periodicity with global significance greater than $3\sigma$. 
Among our samples, some showed quasi-periodic oscillation (QPO) characteristics in their \gr\ light curves.
There are 12 blazars have been reported to have \gr\ QPOs according to Table 2 in \citet{wggapj}, while nearly 30 blazars have been reported to show possible QPOs with high-significance based on \lat\ data so far. 
We note that various analysis methods can be affected by several caveats or effects that may have an impact when analysing time series, and lead to the overestimation of signal significance. 
The caveats remind us of the complexity of the QPO analysis in AGNs, and the importance of correction for trials when computing probabilities. 
\citet{ote+23,ren+23} provided a detailed discussion of some of the caveats.

\subsection{The spectral behavior}
\label{subsec:spectral_behavior}
Variability is one of the main characteristics of blazars, the variability time scale spans from years to hours and even to minutes.
The variability of flux is always accompanied by the variation of spectra.
The correlation between the spectral index and flux has been investigated for individual sources and also for large samples \citep{Fiorucci2004A&A419, Gu2006A&A450, Dai2009MNRAS392, Bonning2012ApJ756, Yuan2017A&A605, Raiteri2017Natur552, Meng2018ApJS237, Xiong2020ApJS247, Safna2020MNRAS498}.
In general, this correlation was mainly discussed at the optical band and demonstrates `bluer-when-brighter (BWB)' behavior for BL Lacs, and shows `redder-when-brighter (RWB)' behavior for FSRQs, except in some special cases e.g., 14 out of 29 Sloan Digital Sky Survey (SDSS) FSRQs show BWB trend \citep{Gu2011A&A528}, 2 out of 40 \textit{Fermi} FSRQs exhibit BWB trend and 7 out of 13 BL Lacs exhibit RWB trend \citep{Zhang2022ApJS259}.
Various models have been proposed to explain blazar optical spectral behavior, shock-in-jet model \citep{Rani2010MNRAS404}, two-components (one variable + one stable) or one synchrotron model \citep{Fiorucci2004A&A419}, the energy injection model \citep{Spada2001MNRAS325, Zhang2002ApJ572}, and the also the vary of beaming effect \citep{Larionov2010A&A510A}.
Recently, \citet{Zhang2022ApJS259} suggests a universal two component-model to interpret these two spectral behaviors, in which the observed optical emission of blazars consists of a stable or less-variable thermal emission component ( $F_{\rm ther}$) primarily coming from the accretion disk, and a highly variable non-thermal emission component ($F_{\rm syn}$) coming from the jet.
The stronger the thermal emission component the bluer the color is, the weaker the thermal emission the redder the color is.

However, the spectral behavior at higher energy bands seems monochrome.
We found a universal BWB trend at $\gamma$-ray band for the TeV blazars in our sample, especially the LBLs and FSRQs showing strong anti-correlation between the photon index and the GeV $\gamma$-ray luminosity.
For the individual sources, \citet{Hayashida2012ApJ754} performed a broadband study of 3C 279 flare and found BWB trend at the X-ray band and $\gamma$-ray bands.
And this BWB trend was found again at the X-ray band for the same source during a phase of increased activity from 2013 December to 2014 April \citep{Hayashida2015ApJ807}.
Moreover, \citet{Aleksic2014A&A569} made multi-frequency observations of PKS 1510-089 in early 2012 and reported a BWB trend at the X-ray band.
\citet{Prince2017ApJ844} studied the long-term light curve of PKS 1510-089 at GeV bands and reported the BWB trend during flares at different campaigns.
There are 14 outbursts/flares of individual TeV blazars that have been analyzed and their spectral behavior has been illustrated in Figure \ref{fig:flux_index}.
11 out of the 14 sources show the BWB trend, according to the insets in Figure \ref{fig:flux_index}, except 4FGL J0221.1+3556, 4FGL J0303.4$-$2407, and 4FGL J1512.8$-$0906.
We suggest this spectral behavior for blazars at the GeV band arises from the same mechanism, which is that the synchrotron-self Compton (SSC) process dominates the GeV emission for these TeV blazars.
Considering the non-thermal electrons that produce the observed inverse Compton emission with an energy distribution
\begin{equation}
    \frac{{\rm d}N}{{\rm d}\gamma} = N_{\rm 0}\, \gamma^{\rm -\alpha}, \,\,\,\,\, \gamma_{\rm min} \leq \gamma \leq \gamma_{\rm max},
\end{equation}
where $\gamma$ is the Lorentz factor of electrons, $\gamma_{\rm min}$ and $\gamma_{\rm max}$ are the minimum and the maximum values of Lorentz factor at the time of particle injection, $N_{\rm 0}$ is related to the total particle density $N_{\rm tot}$ by $N_{\rm 0} = N_{\rm tot} (1-\alpha)/(\gamma_{\rm max}^{1-\alpha} - \gamma_{\rm min}^{1-\alpha})$, $\alpha$ is the electron spectral index.
Then, the SSC emissivity ($j_{\rm ssc}$) is related to the electron spectral index by
\begin{equation}
    j_{\rm ssc} (\epsilon) \sim \epsilon^{-(2+\alpha)/4},
\label{jssc}
\end{equation}
where $\epsilon = h\nu/m_{\rm e}c^{2}$ \citep{Chiang2002ApJ564}.
From equation \ref{jssc}, we can see that the spectral behavior at the GeV band for blazars is mainly determined by the shape of the electron spectrum, which means a harder electron spectrum results in a corresponding harder emission spectrum.
In this case, we can obtain the electron spectral index through the GeV $\gamma$-ray photon index via $-(\Gamma -1) = -(2+\alpha)/4$ for the 11 outbursts/flares and list the results in column (6) of Table \ref{tab:flare}.

\section{Summary} \label{sec:5}

This paper aims to provide detailed GeV variability of the TeV blazars and study the GeV spectral behaviors. 
We performed an analysis using the LAT data across 15 years and offered annual GeV fluxes and corresponding photon spectral indices for the 78 TeV blazars of our sample.
We calculated the detailed monthly flux and corresponding photon spectral index of the 41 bright TeV blazars to further investigate the spectral behavior. 
A series of variability property analyses were conducted on the fractional variability, flux distribution, flare duty cycle, PSDs, and periodic properties.

Our main conclusions are as follows:

(1) We investigated the possible correlation between GeV luminosity and spectral photon index. 
The results suggest a strong correlation between the $\log L_{\rm \gamma}$ and \gr\ photon index for the FSRQs and HBLs, while no correlation for the IBLs.

(2) There are 14 sources out of our sample that show significant flares, of which 6 exhibit a clear sharp peak profile in their 5-day binned light curves. 
4FGL J0303.4$-$2407 and 4FGL J0739.2+0137 show a fast-rise-slow-decay subflare. 
This asymmetry can be related to the particle acceleration mechanism in the jet. 
While 4FGL J1751.5+0938 shows a slow-rise-fast-decay subflare, which may be associated with an efficient cooling process.

(3) We quantified the variability utilizing the fractional variability parameter $F_{\mathrm{var}}$ and the results indicate that the flux of the FSRQs showed significantly stronger variability than that of the BL Lacs. 
As the synchrotron peak frequency decreases, the $F_{\mathrm{var}}$ value generally becomes larger.

(4) We constructed histograms of the observed GeV light curves and fitted them to two different PDFs, a normal distribution and a log-normal distribution. 
The results show that all of the bright sources in this work support a log-normal distribution. 

(5) Our duty cycle results of the monthly-binned light curves show values ranging from 0.0 to 0.36, while the vast majority of the values are in the range of 0.0 to 0.2 except for three blazars.

(6) We found that the periodograms are consistent with a power-law form with the slope index $\beta$ ranging between 0.22$-$1.98. 
Our PSD result at the GeV band is slightly flatter and has a larger range compared with the previous studies.
In addition, 12 blazars in our sample have been reported to have high-significance \gr\ QPOs.

(7) Through checking the spectral behavior, we found 11 out of the 14 sources show a ‘bluer-when-brighter’ trend, which suggests this spectral behavior at the GeV band arises from the mechanism that the synchrotron-self Compton process dominates the GeV emission for these TeV blazars. 

\begin{acknowledgments}

Thanks are given to the reviewer for the constructive comments and suggestions that helped us to make the paper more thorough. This work is supported by the National Natural Science Foundation of China (Grants Nos. 11975072, 11835009, and 11805031), the National SKA Program of China (Grants Nos. 2022SKA0110200 and 2022SKA0110203), the 111 Project (Grant No. B16009), and the China Postdoctoral Science Foundation No. 2023M730523.

\end{acknowledgments}

\bibliography{Gev_vari}
\bibliographystyle{aasjournal}

\end{document}